%% file: draft.tex
\documentclass{ar-1col}
\input{preamble.tex}
\usepackage{amsmath}
\usepackage{graphicx}
\usepackage[colorlinks=true, urlcolor=blue, citecolor=blue, linkcolor=black]{hyperref}
\jname{Annu. Rev. Astron. Astrophys.}
\jvol{55}
\fstpage{343}
\endpage{387}
\jyear{2017}
\doi{10.1146/annurev-astro-091916-055313}
\firstpagenote{Draft version. Posted with permission from the Annual Review of Astronomy and Astrophysics, Volume 55 by Annual Reviews, http://www.annualreviews.org}
 
\begin{document}
\markboth{Bullock \textbullet\ Boylan-Kolchin}{Challenges to the \lcdm\ Paradigm}

\title{Small-Scale Challenges to the $\bm{\Lambda}$CDM Paradigm}

\author{James S. Bullock$^1$ and Michael Boylan-Kolchin$^2$
\affil{$^1$Department of Physics and Astronomy, University of California,
  Irvine, CA 92697, USA; email: bullock@uci.edu} 
\affil{$^2$Department of Astronomy, The University of Texas at Austin, 2515
  Speedway, Stop C1400, Austin, TX 78712, USA; email: mbk@astro.as.utexas.edu}}

\begin{abstract}
  The dark energy plus cold dark matter (\lcdm) cosmological model has been a
  demonstrably successful framework for predicting and explaining the
  large-scale structure of Universe and its evolution with time. Yet on
  length scales smaller than $\sim 1$ Mpc and mass scales smaller than
  $\sim 10^{11} \msun$, the theory faces a number of challenges. For example,
  the observed cores of many dark-matter dominated galaxies are both less dense
  and less cuspy than naively predicted in \lcdm. The number of small galaxies
  and dwarf satellites in the Local Group is also far below the predicted count of
  low-mass dark matter halos and subhalos within similar volumes. These issues
  underlie the most well-documented problems with \lcdm: Cusp/Core, Missing
  Satellites, and Too-Big-to-Fail. The key question is whether a better understanding of baryon physics, dark matter physics, or both will be required to meet these
  challenges. Other anomalies, including the observed planar and orbital
  configurations of Local Group satellites and the tight baryonic/dark matter
  scaling relations obeyed by the galaxy population, have been less thoroughly
  explored in the context of \lcdm\ theory. Future surveys to discover faint,
  distant dwarf galaxies and to precisely measure their masses and density
  structure hold promising avenues for testing possible solutions to the
  small-scale challenges going forward. Observational programs to constrain or discover and characterize the 
the number of truly dark low-mass halos are
  among the most important, and achievable, goals in this field over then next
  decade. These efforts will either further verify the \lcdm\ paradigm or demand
  a substantial revision in our understanding of the nature of dark matter.
\end{abstract}

\maketitle

\tableofcontents

\section{INTRODUCTION}

Astrophysical observations ranging from the scale of the horizon ($\sim$ 15,000
Mpc) to the typical spacing between galaxies ($\sim$ 1 Mpc) are all consistent
with a Universe that was seeded by a nearly scale-invariant fluctuation spectrum
and that is dominated today by dark energy ($\sim 70 \%$) and Cold Dark Matter
($\sim 25\%$), with baryons contributing only $\sim 5\%$ to the energy density
\citep[][]{planck2015,guo16}. This cosmological model has provided a compelling
backbone to galaxy formation theory, a field that is becoming increasingly
successful at reproducing the detailed properties of galaxies, including their
counts, clustering, colors, morphologies, and evolution over time
\citep{vogelsberger14,Schaye15}.  As described in this review, there are
observations below the scale of $\sim 1$ Mpc that have proven more
problematic to understand in the \lcdm\ framework.  It is not yet clear whether
the small-scale issues with \lcdm\ will be accommodated by a better
understanding of astrophysics or dark matter physics, or if they will require a
radical revision of cosmology, but any correct description of our Universe must
look very much like \lcdm\ on large scales.  It is with this in mind that we
discuss the small-scale challenges to the current paradigm. For concreteness, we
assume that the default \lcdm\ cosmology has parameters
$h=H_0/(100 \,\kms \,\mpc^{-1}) = 0.6727$, $\Omega_{m} = 0.3156$,
$\Omega_{\Lambda} = 0.6844$, $\Omega_b = 0.04927$, $\sigma_8 = 0.831$, and
$n_s = 0.9645$ \citep{planck2015}.

Given the scope of this review, we must sacrifice detailed discussions for a
more broad, high-level approach. There are many recent reviews or overview
papers that cover, in more depth, certain aspects of this review. These include \citet{frenk2012}, \citet{peebles2012a}, and \citet{primack2012} on the
historical context of \lcdm\ and some of its basic predictions;
\citet{willman2010} and \citet{mcconnachie2012} on searches for and observed
properties of dwarf galaxies in the Local Group; \citet{feng2010},
\citet{porter2011}, and \citet{strigari2013} on the nature of and searches for
dark matter; \citet{kuhlen2012b} on numerical simulations of cosmological structure
formation; and \cite{brooks2014a}, \citet{weinberg2015} and \citet{del-popolo2017} on small-scale
issues in \lcdm.  Additionally, we will not discuss cosmic acceleration (the
$\Lambda$ in \lcdm) here; that topic is reviewed in \citet{weinberg2013}.
Finally, space does not allow us to address the possibility that the challenges
facing \lcdm\ on small scales reflects a deeper problem in our understanding of
gravity.  We point the reader to reviews by \citet{Milgrom2002}, \citet{famaey2012}, and 
\citet{McGaugh2015}, which compare Modified Newtonian Dynamics (MOND) to \lcdm\
and provide further references on this topic.

\vspace{-0.1cm}
\subsection{Preliminaries: how small is a small galaxy?}
\label{subsec:small_galaxy}
This is a review on small-scale challenges to the \lcdm\ model.  The past
$\sim 12$ years have seen transformative discoveries that have fundamentally
altered our understanding of ``small scales'' -- at least in terms of the
low-luminosity limit of galaxy formation.

Prior to 2004, the smallest galaxy known was Draco, with a stellar mass of
$\mstar \simeq 5 \times 10^5 \msun$.  Today, we know of galaxies 1000 times less
luminous. While essentially all Milky Way satellites discovered before 2004 were
found via visual inspection of photographic plates (with the exceptions of
the Carina and Sagittarius dwarf spheroidal galaxies), the advent of large-area digital sky surveys
with deep exposures and accurate star-galaxy separation algorithms has
revolutionized the search for and discovery of faint stellar systems in the
Milky Way (see \citealt{willman2010} for a review of the search for faint
satellites). The Sloan Digital Sky Survey (SDSS) ushered in this revolution,
doubling the number of known Milky Way satellites in the first five years of
active searches.  The PAndAS survey discovered a similar population of faint
dwarfs around M31 \citep{richardson2011}.  More recently the DES survey has
continued this trend \citep{koposov2015, drlica-wagner2015}.  All told, we know
of $\sim 50$ satellite galaxies of the Milky Way and $\sim 30$ satellites of M31
today \citep[][updated on-line catalog]{mcconnachie2012}, most of which are
fainter than any galaxy known at the turn of the century.  They are also
extremely dark-matter-dominated, with mass-to-light ratios within their stellar
radii exceeding $\sim 1000$ in some cases \citep{walker2009,wolf2010}.

Given this upheaval in our understanding of the faint galaxy frontier over the
last decade or so, it is worth pausing to clarify some naming conventions.  In
what follows, the term ``dwarf'' will refer to galaxies with
$\mstar \lesssim 10^9 \msun$.  We will further subdivide dwarfs into three mass
classes: Bright Dwarfs ($\mstar \approx 10^{7-9} \msun$), Classical Dwarfs
($\mstar \approx 10^{5-7} \msun$), and Ultra-faint Dwarfs
($\mstar \approx 10^{2-5} \msun$).  Note that another common classification for
dwarf galaxies is between dwarf spheroidals (dSphs) and dwarf drregulars (dIrrs).
Dwarfs with gas and ongoing star formation are usually labeled dIrr.  The term
dSph is reserved for dwarfs that lack gas and have no ongoing star formation.
Note that the vast majority of field dwarfs (meaning that they are not
satellites) are dIrrs. Most dSph galaxies are satellites of larger systems.

\begin{figure*}[th!]
\includegraphics[width=\textwidth]{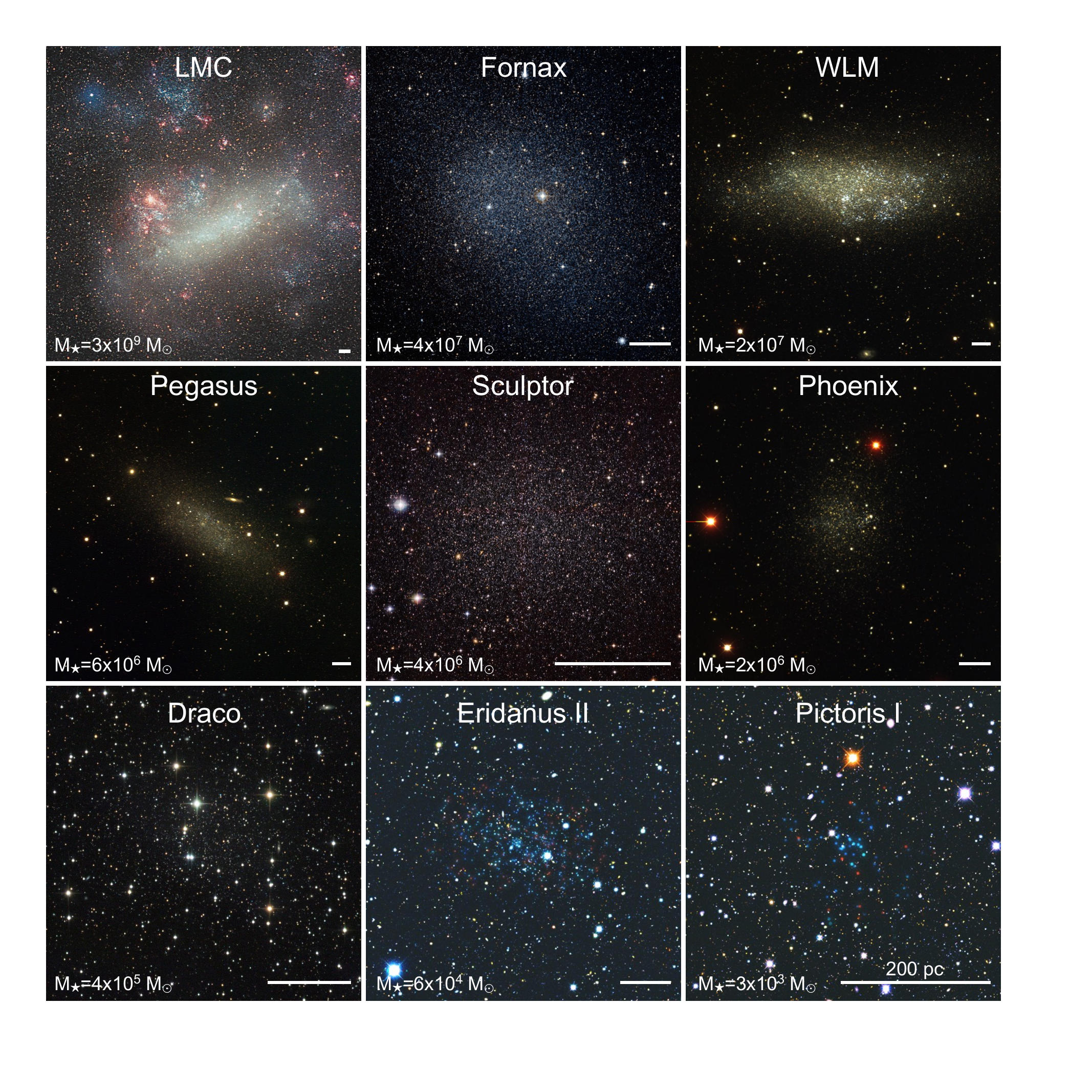}
\caption{\textit{Approaching the threshold of galaxy formation.} Shown
  are images of dwarf galaxies spanning six orders of magnitude in
  stellar mass. In each panel, the dwarf's stellar mass is listed in the lower-left corner and a scale bar corresponding to 200 pc is shown in the lower-right corner. The LMC, WLM,
  and Pegasus are dwarf irregular (dIrr) galaxies that have gas and ongoing
  star formation.  The remaining six galaxies shown are gas-free dwarf spheroidal (dSph) galaxies and are not currently forming stars. The faintest galaxies shown here are only detectable in limited volumes around the Milky Way; future surveys may reveal many more such galaxies at greater distances. Image credits: Eckhard Slawik (LMC); ESO/Digitized Sky Survey 2 (Fornax); \citeauthor{massey2007} (2007; WLM, Pegasus, Phoenix); ESO (Sculptor); Mischa Schirmer (Draco), Vasily Belokurov and Sergey Koposov (Eridanus II, Pictoris I).  \vspace{2cm}
\label{fig:dwarfs}
}
\end{figure*}

Figure \ref{fig:dwarfs} illustrates the morphological differences among galaxies
that span these stellar mass ranges.  From top to bottom we see three dwarfs
each that roughly correspond to Bright, Classical, and Ultra-faint Dwarfs,
respectively.

\begin{summary}[ADOPTED DWARF GALAXY NAMING CONVENTION]
  \noindent {\bf Bright Dwarfs:} $\mstar \approx 10^{7-9} \msun$ \\ -- the faint
  galaxy completeness limit for field galaxy surveys

\bigskip
\noindent {\bf Classical Dwarfs:} $\mstar \approx 10^{5-7} \msun$ \\ -- the
faintest galaxies known prior to SDSS

\bigskip
\noindent {\bf Ultra-faint Dwarfs:} $\mstar \approx 10^{2-5} \msun$ \\ --
detected within limited volumes around M31 and the Milky Way
\end{summary}

With these definitions in hand, we move to the cosmological model within which
we aim to explain the counts, stellar masses, and dark matter content of these
dwarfs.

\subsection{Overview of the \lcdm\ model}
\label{subsec:lcdm}
The \lcdm\ model of cosmology is the culmination of century of work on the
physics of structure formation within the framework of general relativity. It
also indicates the confluence of particle physics and astrophysics over the past
four decades: the particle nature of dark matter directly determines essential
properties of non-linear cosmological structure. While the \lcdm\ model is
phenomenological at present -- the actual physics of dark matter and dark energy
remain as major theoretical issues -- it is highly successful at explaining the
large-scale structure of the Universe and basic properties of galaxies that form
within dark matter halos.

\begin{figure}
\centering
\includegraphics[width=0.95\textwidth]{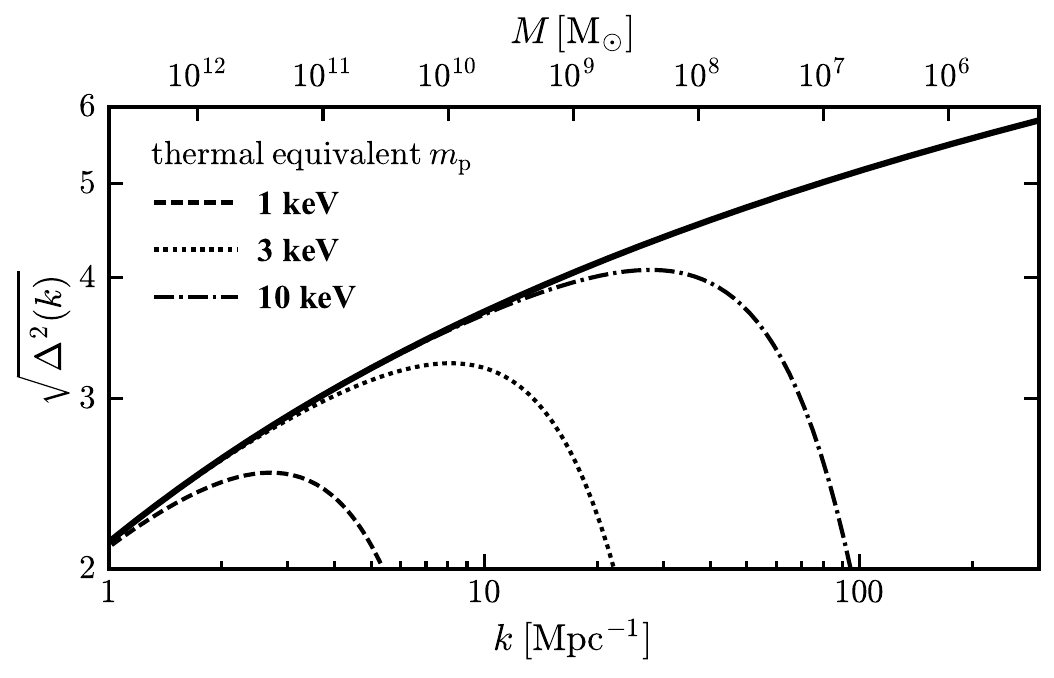} \\
\includegraphics[width=0.95\textwidth]{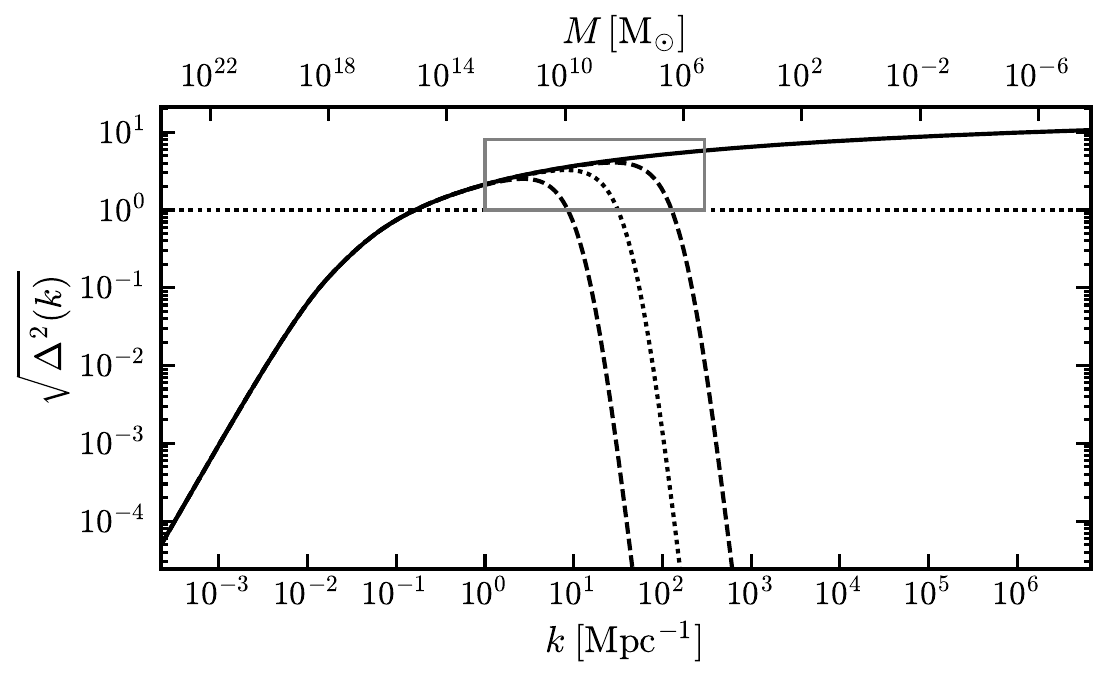}
\caption{The \lcdm\ dimensionless power spectrum (solid lines, Equation \ref{eq:5}) plotted
  as a function of linear wavenumber $k$ (bottom axis) and corresponding linear
  mass $M_l$ (top axis). The bottom panel spans all physical scales relevant for
  standard CDM models, from the particle horizon to the
  free-streaming scale for dark matter composed of standard 100 GeV WIMPs on the
  far right. The top panel zooms in on the scales of interest for this review,
  marked with a rectangle in the bottom panel. Known
  dwarf galaxies are consistent with occupying a relatively narrow 2 decade
  range of this parameter space -- $10^{9}-10^{11}\,\msun$ -- even though dwarf
  galaxies span approximately 7 decades in stellar mass. The effect of WDM models on the
  power spectrum is illustrated by the dashed, dotted, and dash-dotted lines, which
  map to the (thermal) WDM particle masses listed.  See Section \ref{subsubsec:linear} for a discussion
  of power suppression in WDM.}
\label{fig:pofk}
\end{figure}

In the \lcdm\ model, cosmic structure is seeded by primordial adiabatic
fluctuations and grows by gravitational instability in an expanding
background. The primordial power spectrum as a function of wavenumber $k$ is
nearly scale-invariant\footnote{Recent measurements find $n = 0.968 \pm 0.006$
  \citep{planck2015}, i.e., small but statistically different from true scale
  invariance.}, $P(k)\propto k^n$ with $n\simeq 1$.
Scales that re-enter the horizon when the Universe is radiation-dominated grow
extremely slowly until the epoch of matter domination, leaving a scale-dependent
suppression of the primordial power spectrum that goes as $k^{-4}$ at large
$k$. This suppression of power is encapsulated by the ``transfer function" $T(k)$, which is defined as the ratio of amplitude of a density perturbation in the post-recombination era to its primordial value as a function of perturbation wavenumber $k$.
This processed power
spectrum is the input for structure formation calculations; the dimensionless
processed power spectrum, defined by
\begin{equation}
  \label{eq:5}
  \Delta^2(k, a)=\frac{k^3}{2\pi^2}\,P(k)\,T^2(k)\,d^2(a)\,,
\end{equation}
therefore rises as $k^4$ for scales larger than the comoving horizon at
matter-radiation equality (corresponding to $k=0.008\,\mpc^{-1}$) and is
approximately independent of $k$ for scales that re-enter the horizon well
before matter-radiation equality.   Here, $d(a)$ is the linear growth function,
normalized to unity at $a=1$. The processed $z=0$ ($a=1$)  linear power spectrum for \lcdm~ is shown by the solid
line in Figure \ref{fig:pofk}.  The asymptotic shape behavior is most easily seen in the bottom panel, which spans the
wave number range of cosmological interest. 
For a more complete discussion of primordial fluctuations and the processed power spectrum 
we recommend that readers consult \citet{mo2010}.

It is useful to associate each wavenumber with a mass scale set by its
characteristic length $r_l=\lambda/2=\pi/k$.  In the early Universe, when
$\delta \ll 1$, the total amount of matter contained within a sphere of comoving
Lagrangian radius $r_l$ at $z=0$ is

\begin{eqnarray}
  \label{eq:mlin}
  M_l&=&\frac{4\,\pi}{3}\,r_l^3\,\rho_{\rm m} = \frac{\Omega_{\rm m}\,H_0^2}{2\,G}\,r_l^3\,\\
  &=&1.71 \times 10^{11}\,\msun \left(\frac{\Omega_{\rm m}}{0.3}\right)
         \left(\frac{h}{0.7}\right)^2\,\left(\frac{r_l}{1\,\mpc}\right)^3\, .
\end{eqnarray}

The mapping between wave number and mass scale is illustrated by the top and bottom axis in Figure \ref{fig:pofk}. 
The processed linear power spectrum for \lcdm\ shown in the bottom panel (solid line)
spans the horizon scale to a typical mass cutoff scale for the most common cold
dark matter candidate ($\sim 10^{-6} \msun$; see discussion in
Section~\ref{subsec:particle_physics}).  A line at $\Delta = 1$ is plotted for
reference, showing that fluctuations born on comoving length scales smaller
than $r_l \approx 10\,\hmpc \approx 14 \,\mpc$ have gone non-linear today.  The
top panel is zoomed in on the small scales of relevance for this review (which
we define more precisely below).  Typical regions on these scales have collapsed
into virialized objects today.  These collapsed objects -- dark matter halos --
are the sites of galaxy formation.

\subsection{Dark matter halos}
\label{subsec:lcdm_small}

\subsubsection{Global properties}
Soon after overdense regions of the Universe become non-linear, they stop expanding, turn around, and collapse, converting potential energy into kinetic energy in the process.  
The result is virialized dark matter halos with masses given by
\begin{equation}
  \label{eq:mvir}
  \mvir=\frac{4\,\pi}{3}\,\rvir^3\,\Delta\,\rho_{\rm m}\,,
\end{equation}
where $\Delta \sim 300$ is the virial over-density parameter, defined here
relative to the background matter density.  As discussed below, the value of
$\mvir$ is ultimately a definition that requires some way of defining a halo's
outer edge ($\rvir$).  This is done via a choice for $\Delta$.  The numerical
value for $\Delta$ is often chosen to match the over-density one predicts for a
virialized dark matter region that has undergone an idealized spherical collapse
\citep[][]{bryan1998}, and we will follow that convention here.  Note that given
a virial mass $\mvir$, the virial radius, $\rvir$, is uniquely defined by
Equation \ref{eq:mvir}.  Similarly, the virial {\em velocity}
\begin{equation}
\vvir \equiv \sqrt{\frac{G \mvir}{\rvir}},
\end{equation}
is also uniquely defined.  The parameters $\mvir$, $\rvir$, and $\vvir$ are
equivalent mass labels -- any one determines the other two, given a specified over-density parameter $\Delta$.

\begin{marginnote}[]
\entry{Galaxy Clusters}{$\mvir\approx 10^{15} \msun$ $\vvir \approx 1000 \, \kms$}
\entry{Milky Way}{$\mvir\approx 10^{12} \msun$ $\vvir \approx 100 \, \kms$}
\entry{Smallest Dwarfs}{$\mvir\approx 10^{9} \msun$ $\vvir \approx 10 \, \kms$}
\end{marginnote}

One nice implication of Equation \ref{eq:mvir} is that a present-day object with
virial mass $\mvir$ can be associated directly with a linear perturbation with
mass $M_l$. Equating the two gives
\begin{equation}
  \label{eq:rvir}
  \rvir=0.15 \,  \left(\frac{\Delta}{300}\right)^{-1/3}\,r_l\,.
\end{equation}
We see that a collapsed halo of size $\rvir$ is approximately 7 times smaller in
physical dimension than the comoving linear scale associated with that mass
today.

With this in mind, Equations 3-6 allow us to self-consistently define ``small
scales" for both the linear power spectrum and collapsed objects:
$M \lesssim 10^{11} \, \msun$.  As we will discuss, potential problems
associated with galaxies inhabiting halos with $\vvir \simeq 50 \, \kms$ may
point to a power spectrum that is non-CDM-like at scales $r_l \la 1 \,\mpc$.

\begin{summary}[WE DEFINE ``SMALL SCALES'' AS THOSE SMALLER THAN:]
\vspace{-0.25cm}
\begin{equation*}
  \label{eq:4}
  M \approx 10^{11}\,\msun \leftrightarrow k \approx 3\,\mpc^{-1} 
  \leftrightarrow r_l \approx 1\,\mpc 
  \leftrightarrow \rvir \approx 150 \,\kpc 
  \leftrightarrow \vvir \approx 50\,\kms\,.  
\end{equation*}
\end{summary}

As alluded to above, a common point of confusion is that the halo mass
definition is subject to the assumed value of $\Delta$, which can vary by a
factor of $\sim 3$ depending on the author.  For the spherical collapse
definition, $\Delta \simeq 333$ at $z=0$ (for our fiducial cosmology) and
asymptotes to $\Delta = 178$ at high redshift \citep[][]{bryan1998}.  Another common
choice is a fixed $\Delta = 200$ at all $z$ (often labeled $M_{200m}$ in the
literature).  Finally, some authors prefer to define the virial overdensity as
$200$ times the critical density, which, according to Equation \ref{eq:mvir}
would mean $\Delta(z) = 200 \rho_{c}(z)/\rho_m(z)$.  Such a mass is commonly
labeled ``$M_{200}$'' in the literature.  For most purposes (e.g., counting
halos), the precise choice does not matter, as long as one is consistent with the
definition of halo mass throughout an analysis: every halo has the same center,
but its outer radius (and mass contained within that radius) shifts depending on
the definition.  In what follows, we use the spherical collapse definition
($\Delta = 333$ at $z=0$) and adhere to the convention of labeling that mass
``$\mvir$".

Before moving on, we note that it is also possible (and perhaps even preferable)
to give a halo a ``mass" label that is directly tied to a physical feature
associated with a collapsed dark matter object rather than simply adopting a
$\Delta$.  \citet{more2015} have advocated the use of a ``splash-back" radius ,
where the density profile shows a sharp break (this typically occurs at
$\sim 2 \rvir$).  Another common choice is to tag halos based not on a mass but
on $\vmax$, which is the peak value of the circular velocity $V_c(r) = \sqrt{G M(<r)/r}$
as one steps out from the halo center.  For any individual halo, the value of
$\vmax$ ($\gtrsim \vvir$) is linked to the internal mass profile or density
profile of the system, which is the subject of the next subsection.  As
discussed below, the ratio $\vmax/\vvir$ increases as the halo mass decreases.

\begin{textbox}[ht]\section{ROBUST PREDICTIONS FROM CDM-ONLY SIMULATIONS}
  A defining characteristic of CDM-based hierarchical structure formation is
  that the smallest scales collapse first -- a fact that arises directly from
  the shape of the power spectrum (Figure 1) and that lies at the heart of many
  robust predictions for the counts and structure of dark matter halos today.
  As discussed below, baryonic processes can alter these predictions to various
  degrees, but pure dark matter simulations have provided a well-defined set of
  basic predictions used to benchmark the theory.

\subsection{The dark matter profiles of individual halos are cuspy and dense [Figure \ref{fig:nfw}]}
The density profiles of individual \lcdm\ halos increase steadily towards small
radii, with an overall normalization and detailed shape that reflects the halo's
mass assembly.  At fixed mass, early-forming halos tend to be denser than
later-forming halos.  As with the mass function, both the shape {\em and}
normalization of dark matter halo density structure is predicted by \lcdm, with
a well-quantified prediction for the scatter in halo concentration at fixed
mass.

\subsection{There are many more small halos than large ones [Figure \ref{fig:massfunc}]}
The comoving number density of dark matter halos rises steeply towards small
masses, $dn/dM \propto M^{\alpha}$ with $\alpha \simeq -1.9$.  At large halo
masses, counts fall off exponentially above the mass scale that is just going
non-linear today.  Importantly, both the shape and normalization of the mass
function is robustly predicted by the theory.

\subsection{Substructure is abundant and almost self-similar [Figure \ref{fig:mf}]} 
Dark matter halos are filled with substructure, with a mass function that rises
as $dN/dm \propto m^{\alpha_s}$ with $\alpha_s \simeq -1.8$ down to the low-mass
free-streaming scale ($m \ll 1 \msun$ for canonical models).  Substructure
reflects the high-density cores of smaller merging halos that survive the
hierarchical assembly process.  Substructure counts are nearly self-similar with
host mass, with the most massive subhalos seen at
$m_{\rm max} \sim 0.2 M_{\rm host}$.
  \end{textbox}

\subsubsection{Abundance}
In principle, the mapping between the initial spectrum of density fluctuations
at $z \rightarrow \infty$ and the mass spectrum of collapsed (virialized) dark matter
halos at later times could be extremely complicated: as a given scale becomes
non-linear, it could affect the collapse of nearby regions or larger scales. In
practice, however, the mass spectrum of dark matter halos can be modeled
remarkably well with a few simple assumptions. The first of these was taken by
\citet{press1974}, who assumed that the mass spectrum of collapsed objects could
be calculated by extrapolating the overdensity field using linear theory even
into the highly non-linear regime and using a spherical collapse model
\citep{gunn1972}. In the Press-Schechter model, the dark matter halo mass
function -- the abundance of dark matter halos per unit mass per unit volume at
redshift $z$, often written as $n(M,z)$ -- depends only on the rms amplitude of
the linear dark matter power spectrum, smoothed using a spherical tophat filter
in real space and extrapolated to redshift $z$ using linear theory. Subsequent
work has put this formalism on more rigorous mathematical footing
\citep{bond1991, cole1991, sheth2001}, and this extended Press-Schechter (EPS) theory
yields abundances of dark matter halos that are perhaps surprisingly accurate
(see \citealt{zentner2007} for a comprehensive review of EPS theory). This accuracy is tested
through comparisons with large-scale numerical simulations.

Simulations and EPS theory both find a universal form for $n(M,z)$: the comoving
number density of dark matter halos is a power law with log slope of
$\alpha \simeq -1.9$ for $M \ll M^*$ and is exponentially suppressed for
$M \gg M^*$, where $M^* = M^*(z)$ is the characteristic mass of fluctuations going
non-linear at the redshift $z$ of interest\footnote{The black line in Figure \ref{fig:massfunc} illustrates the
mass function of \lcdm\ dark matter halos.}.  Importantly, given an initial power spectrum of density
fluctuations, it is possible to make highly accurate predictions within \lcdm\
for the abundance, clustering, and merger rates of dark matter halos at any
cosmic epoch.

\subsubsection{Internal structure}
\citet{dubinski1991} were the first to use $N$-body simulations to show that the
internal structure of a CDM dark matter halo does not follow a simple power-law,
but rather bends from a steep outer profile to a mild inner cusp obeying
$\rho(r) \sim 1/r$ at small radii.  More than twenty years later, simulations
have progressed to the point that we now have a fairly robust understanding of
the structure of \lcdm\ halos and the important factors that govern halo-to-halo
variance \citep[e.g.,][]{navarro2010,diemer2015,klypin2016}, at least for dark-matter-only simulations.
 
To first approximation, dark matter halo profiles can be described by a nearly
universal form over all masses, with a steep fall-off at large radii
transitioning to mildly divergent cusp towards the center.  A common way to
characterize this is via the NFW functional form \citep{navarro1997}, which
provides a good (but not perfect) description dark matter profiles:
\begin{equation}
\label{eq:nfw}
\rho(r) = \frac{4 \rho_{-2}}{(r/r_{-2})(1+r/r_{-2})^2}.
\end{equation}
Here, $r_{-2}$ is a characteristic radius where the log-slope of the density
profile is $-2$, marking a transition point from the inner $1/r$ cusp to an
outer $1/r^3$ profile.  The second parameter, $\rho_{-2}$, sets the value of
$\rho(r)$ at $r=r_{-2}$.  In practice, dark matter halos are better described
the three-parameter \citet{Einasto65} profile \citep{Navarro2004,Gao2008}.
However, for the small halos of most concern for this review, NFW fits do almost
as well as Einasto in describing the density profiles of halos in
simulations \citep{Dutton2014}.  Given that the NFW form is slightly simpler, we
have opted to adopt this approximation for illustrative purposes in this review.

\begin{figure}
\centering
\includegraphics[width=\linewidth]{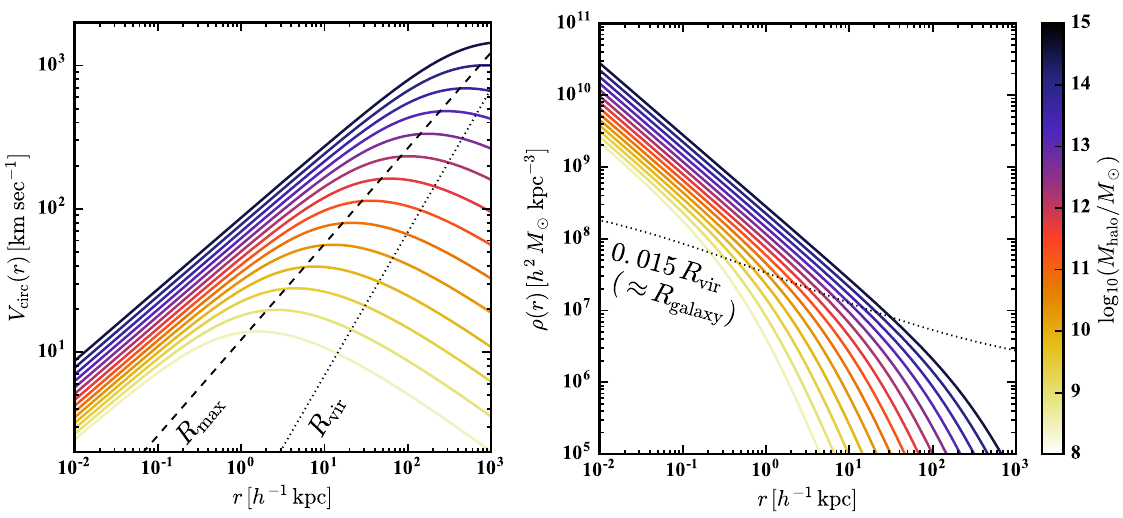}
\caption{{\em Right:} The density profiles of median NFW dark matter halos at
  $z=0$ with masses that span galaxy clusters ($\mvir = 10^{15} \msun$, black)
  to the approximate HI cooling threshold that is expected to correspond to the
  smallest dwarf galaxies ($\mvir \approx 10^8 \msun$, yellow).  The lines are
  color coded by halo virial mass according to the bar on the right and are
  separated in mass by 0.5 dex.  We see that (in the median) massive halos are
  denser than low-mass halos at a fixed {\em physical} radius.  However, at a
  fixed small fraction of the virial radius, smaller halos are typically
  slightly denser than larger halos, reflecting the concentration-mass relation.
  This is demonstrated by the dotted line which connects $\rho(r)$ evaluated at
  $r = \epsilon \rvir$ for halos over a range of masses.  We have chosen
  $\epsilon = 0.015$ because this value provides a good match to observed galaxy
  half-light radii over a wide range of galaxy luminosities under the assumption
  that galaxies occupy halos according to abundance matching (see Section
  \ref{sec:AM} and Figure \ref{fig:AM}).  Interestingly, the characteristic dark
  matter density at this `galaxy radius' increases only by a factor of $\sim 6$
  over almost seven orders of magnitude in halo virial mass.  {\em Left:} The
  equivalent circular velocity curves $V_c(r) \equiv \sqrt{G M(<r)/r}$ for the
  same halos plotted on the right.  The dashed line connects the radius $\rmax$
  where the circular velocity is maximum ($\vmax$) for each rotation curve.  The
  dotted line tracks the $\rvir$ -- $\vvir$ relation.  The ratio $\rmax/\rvir$
  decreases towards smaller halos, reflecting the mass-concentration
  relation. The ratio $\vmax/\vvir$ also increases with decreasing
  concentration.  }
\label{fig:nfw}
\end{figure}

As Equation \ref{eq:nfw} makes clear, two parameters (e.g., $\rho_{-2}$
and $r_{-2}$) are required to determine a halo's NFW density profile.  For a fixed
halo mass $\mvir$ (which fixes $\rvir$), the second parameter is often expressed
as the halo concentration: $c = \rvir/r_{-2}$.  Together, a $\mvir-c$
combination completely specifies the profile.  In the median, and over the mass
and redshift regime of interest to this review, halo concentrations increase
with decreasing mass and redshift: $c \propto \mvir^{-a} \, (1+z)^{-1}$, with
$a \simeq 0.1$ \citep[][]{bullock2001}.  Though halo concentration correlates with halo mass, there is significant scatter ($\sim 0.1$ dex) about the median at fixed $\mvir$
\citep{Jing2000,bullock2001}. Some fraction of this scatter is driven by the
variation in halo mass accretion history \citep{wechsler2002,ludlow2016}, with
early-forming halos having higher concentrations at fixed final virial mass.

The dependence of halo profile on a mass-dependent concentration parameter and the correlation between formation time and concentration at fixed virial mass are caused by the hierarchical build-up of halos in \lcdm: low-mass halos assemble earlier, when the mean density of the Universe is higher, and therefore have higher concentrations than high-mass halos (e.g., \citealt{{navarro1997,wechsler2002}}). At the very smallest masses, the
concentration-mass relation likely flattens, reflecting the shape of the
dimensionless power spectrum \citep[see our Figure 1 and the discussion
in][]{ludlow2016}; at the highest masses and redshifts, characteristic of very
rare peaks, the trend seems to reverse \citep[$a < 0$;][]{klypin2016}.

The right panel of Figure \ref{fig:nfw} summarizes the median NFW density
profiles for $z=0$ halos with masses that span those of large galaxy clusters
($\mvir = 10^{15} \msun$) to those of the smallest dwarf galaxies
($\mvir = 10^8 \msun$).  We assume the $c-\mvir$ relation from \citet{klypin2016}.
These profiles are plotted in physical units (unscaled to the virial radius) in
order to emphasize that higher mass halos are denser at every radius than lower
mass halos (at least in the median).  However, at a fixed small fraction of the
virial radius, small halos are slightly denser than larger ones.  This is a
result of the concentration-mass relation. Under the ansatz of
abundance matching (Section \ref{sec:AM}, Figure \ref{fig:AM}), galaxy sizes
(half-mass radii) track a fixed fraction of their host halo virial radius:
$r_{\rm gal} \simeq 0.015 \rvir$ \citep{kravtsov2013}.  This relation is plotted
as a dotted line such that the dotted line intersects each solid line at that
$r = 0.015\, \rvir$, where $\rvir$ is that particular halo's virial radius.  We
see that small halos are slightly denser at the typical radii of the galaxies
they host than are larger halos.  Interestingly, however, the density range is
remarkably small, with a local density of dark matter increasing by only a
factor of $\sim 6$ over the full mass range of halos that are expected to host
galaxies, from the smallest dwarfs to the largest cD galaxies in the universe.

On the left we show the same halos, now presented in terms of the implied
circular velocity curves: $V_c \equiv \sqrt{GM(<r)/r}$.  The dotted line in left
panel intersects $\vvir$ at $\rvir$ for each value of $\mvir$.  The dashed line does
the same for $\vmax$ and its corresponding radius $\rmax$.  Higher mass systems,
with lower concentrations, typically have $\vmax \simeq \vvir$, but for smaller
halos the ratio is noticeably different than one and can be as large as
$\sim 1.5$ for high-concentration outliers.  Note also that the lowest mass
halos have $\rmax \ll \rvir$ and thus it is the value of $\vmax$ (rather than
$\vvir$) that is more closely linked to the observable ``flat" region of a
galaxy rotation curve.  For our ``small-scale" mass of $\mvir = 10^{11} \msun$,
typically $\vmax \simeq 1.2\, \vvir \simeq 60 \, \kms$.

\begin{figure}[t!]
\centering
\includegraphics[width=0.7\textwidth]{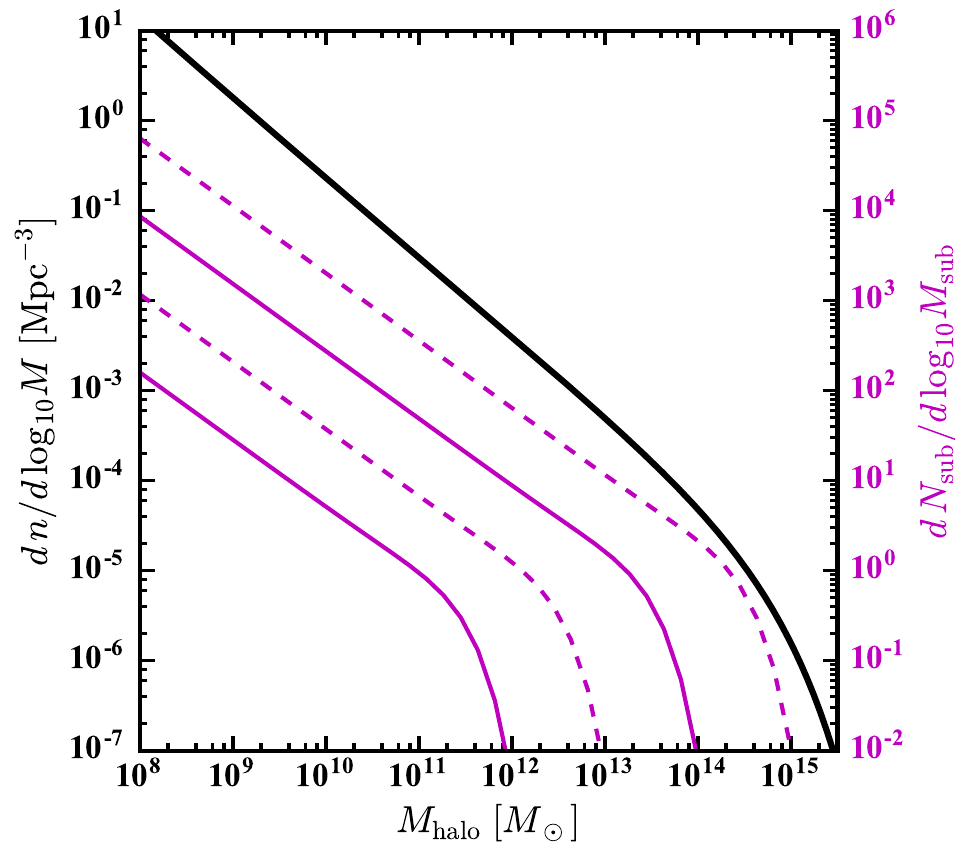}
\caption{Steep mass functions. The black solid line shows the $z=0$ dark matter
  halo mass function ($M_{\rm halo} = \mvir$) for the full population of halos
  in the universe as approximated by \citet{sheth2001}.  For comparison, the
  magenta lines show the subhalo mass functions at $z=0$ (defined as
  $M_{\rm halo} = M_{\rm sub} = \mpeak$, see text) at the same redshift for host halos at four
  characteristic masses ($\mvir = 10^{12}, 10^{13}, 10^{14},$ and
  $10^{15} \msun$) with units given along the right-hand axis.  Note that the
  subhalo mass functions are almost self-similar with host mass, roughly
  shifting to the right by $10\times$ for every decade increase in host mass.
  The low-mass slope of subhalo mass function is similar than the field halo
  mass function.  Both field and subhalo mass functions are expected to rise
  steadily to the cutoff scale of the power spectrum, which for fiducial CDM
  scenarios is $ \ll 1 \msun$.}
\label{fig:massfunc}
\end{figure}

\subsection{Dark matter substructure}

It was only just before the turn of the century that $N$-body simulations set
within a cosmological CDM framework were able to robustly resolve the
substructure {\em within} individual dark matter halos
\citep{Ghigna1998,klypin1999a}.  It soon became clear that the dense centers of
small halos are able to survive the hierarchical merging process: dark matter
halos should be filled with substructure.  Indeed, subhalo counts are nearly self-similar with
host halo mass. This was seen as welcome news for
cluster-mass halos, as the substructure could be easily identified with cluster
galaxies.  However, as we will discuss in the next section, the fact that Milky-Way-size halos are filled with substructure is less clearly consistent with what
we see around the Galaxy. 

Quantifying subhalo counts, however, is not so straightforward.  Counting by
mass is tricky because the definition of ``mass" for an extended distribution
orbiting within a collapsed halo is even more fraught with subjective decisions
than virial mass.  When a small halo is accreted into a large one, mass is
preferentially stripped from the outside.  Typically, the standard virial
overdensity ``edge" is subsumed by the ambient host halo.  One option is to
compute the mass that is bound to the subhalo, but even these masses vary from
halo finder to halo finder.  The value of a subhalo's $\vmax$ is better defined,
and often serves as a good tag for quantifying halos.

Another option is to tag bound subhalos using the maximum virial mass that the halos had
at the time they were first accreted\footnote{This maximum mass is similar to the virial mass at the time of accretion,
though infalling halos can begin losing mass prior to first crossing the virial radius.} onto a host, $\mpeak$.  This is a
useful option because stars in a central galaxy belonging to a halo at accretion
will be more tightly bound than the dark matter.  The resultant satellite's
stellar mass is most certainly more closely related to $\mpeak$ than the
bound dark matter mass that remains at $z=0$.  Moreover, the subsequent mass
loss (and even $\vmax$ evolution) could change depending on the baryonic content
of the {\em host} because of tidal heating and other dynamical effects
\citep{DOnghia2010}.  For these reasons, we adopt $M_{\rm sub} = \mpeak$ for
illustrative purposes here.

The magenta lines in Figure \ref{fig:massfunc} show the median subhalo mass
functions ($M_{\rm sub} = \mpeak$) for four characteristic host halo masses
($\mvir = 10^{12-14} \msun$) according to the results of
\citet{rodriguez-puebla2016}.  These lines are normalized to the right-hand
vertical axis.  Subhalos are counted only if they exist within the virial radius
of the host, which means the counting volume increases as
$\propto \mvir \propto \rvir^3$ for these four lines.  For comparison, the black
line (normalized to the left vertical axis) shows the global halo mass function
\citep[as estimated via the fitting function from][]{sheth2001}.  The subhalo
mass function rises with a similar (though slightly shallower) slope as the field halo mass function and is
also roughly self-similar in host halo mass.

\subsection{Linking dark matter halos to galaxies}
\label{sec:AM}

\begin{figure}[t]
\includegraphics[width=\textwidth]{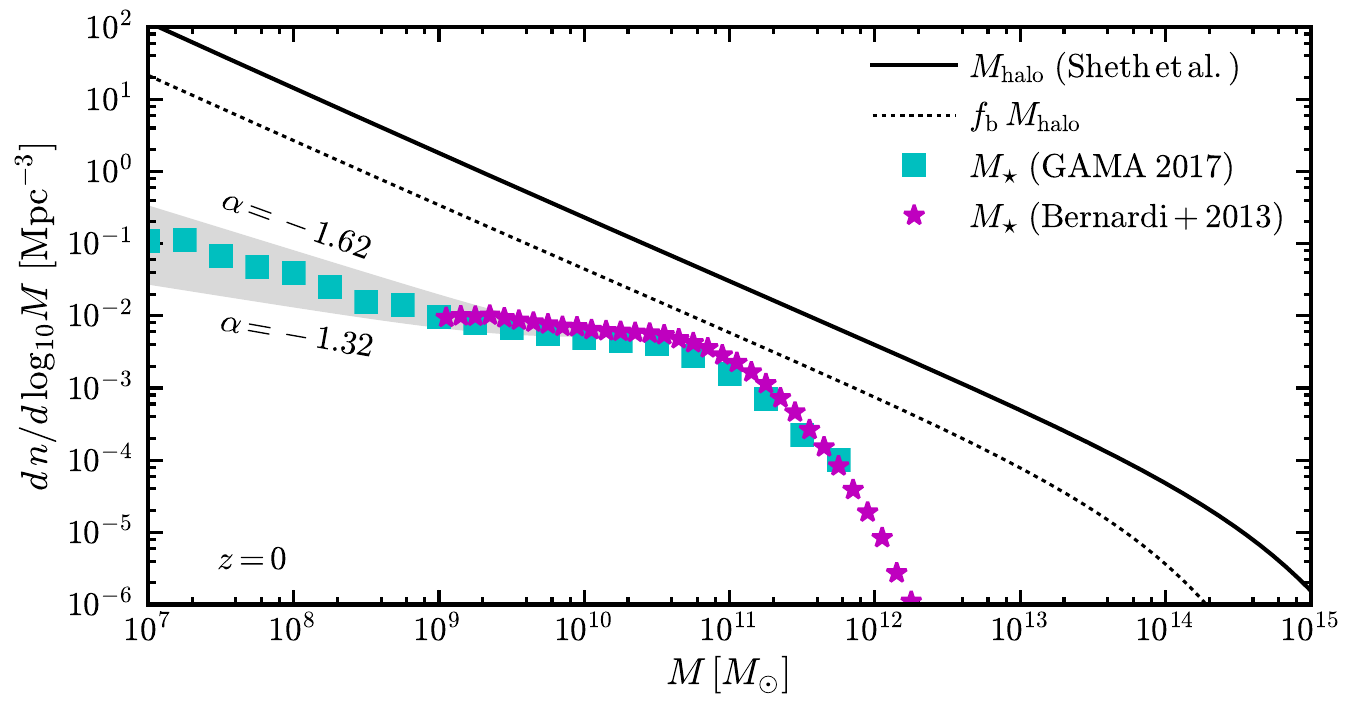}
\caption{The thick black line shows the global dark matter mass function.  The
  dotted line is shifted to the left by the cosmic baryon fraction for each halo
  $\mvir \rightarrow f_b \mvir$.  This is compared to the observed stellar mass
  function of galaxies from \citet[][magenta stars]{bernardi2013} and
 \citeauthor{wright2017} (2017; cyan squares).  The shaded bands 
  demonstrate a range of faint-end slopes
  $\alpha_g = -1.62$ to $-1.32$.  This range of power laws will
  produce   dramatic differences at the scales of the classical Milky Way satellites
  ($\mstar \simeq 10^{5-7} \msun$). Pushing large sky surveys down below
  $10^6 \msun$ in stellar mass, where the differences between the power law
  range shown would exceed a factor of ten, would provide a powerful constraint on our
  understanding of the low-mass behavior.  Until then, this mass regime can only
  be explored with without large completeness corrections in vicinity of the
  Milky Way.  }
\label{fig:mf}
\end{figure}

How do we associate dark matter halos with galaxies? One simple approximation is
to assume that each halo is allotted its cosmic share of baryons
$f_b = \Omega_b/\Omega_m \approx 0.15$ and that those baryons are converted to
stars with some constant efficiency $\epsilon_\star$:
$M_{\star} = \epsilon_\star \, f_b \, \mvir$. Unfortunately, as shown in Figure
\ref{fig:mf}, this simple approximation fails miserably.  Galaxy stellar masses
do not scale linearly with halo mass; the relationship is much more
complicated.  Indeed, the goal of forward modeling galaxy formation from known
physics within the \lcdm\ framework is an entire field of its own (galaxy
evolution; \citealt{somerville2015}).  Though galaxy formation theory has
progressed significantly in the last several decades, many problems remain unsolved.

Other than forward modeling galaxy formation, there are two common approaches
that give an independent assessment of how galaxies relate to dark matter halos.
The first involves matching the observed volume density of galaxies of a given
stellar mass (or other observable such as luminosity, velocity width, or baryon
mass) to the predicted abundance of halos of a given virial mass.  The second
way is to measure the mass of the galaxy directly and to infer the dark matter
halo properties based on this dynamical estimator.

\subsubsection{Abundance matching} As illustrated in Figure \ref{fig:mf}, the
predicted mass function of collapsed dark matter halos has a considerably
different normalization and shape than the observed stellar mass function of
galaxies. The difference grows dramatically at both large and small masses, with
a maximum efficiency of $\epsilon_\star \simeq 0.2$ at the stellar mass scale of
the Milky Way ($\mstar \approx 10^{10.75} \msun$).  This basic mismatch in shape
has been understood since the earliest galaxy formation models set within the
dark matter paradigm \citep{white1978} and is generally recognized as one of the
primary constraints on feedback-regulated galaxy formation
\citep{white1991,benson2003,somerville2015}.

\begin{figure}[t]
\includegraphics[width=0.8\textwidth]{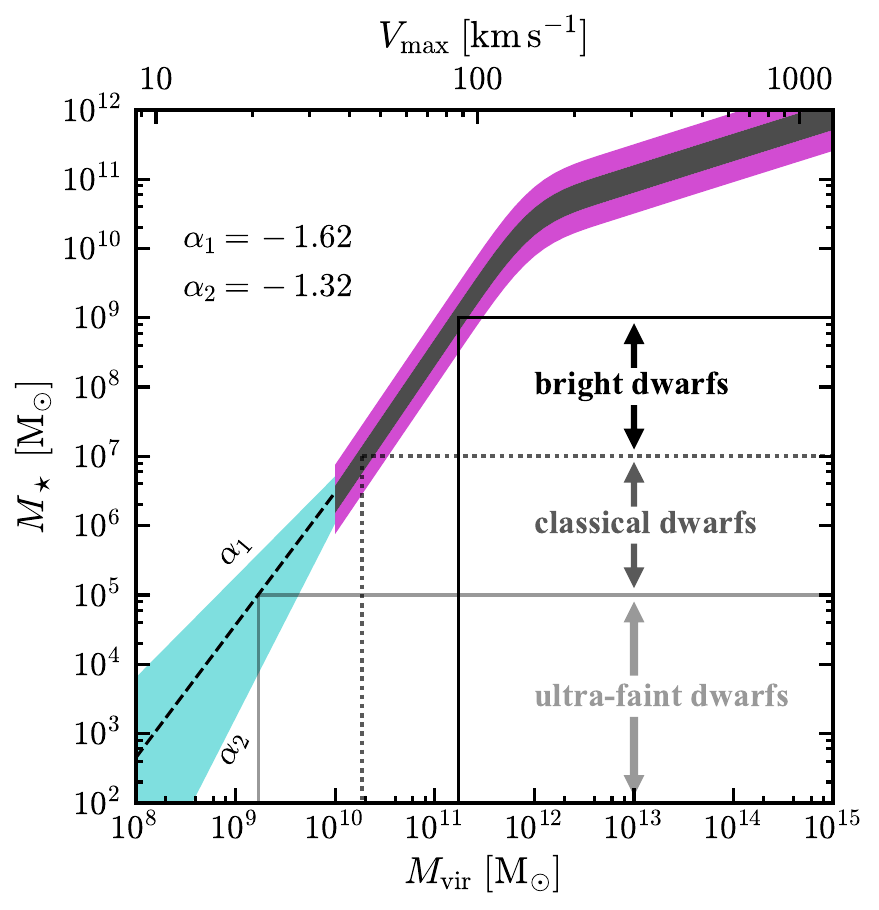}
\caption{Abundance matching relation from Behroozi et al.~(in preparation).
  Gray (magenta) shows a scatter of 0.2 (0.5) dex about the median relation.
  The dashed line is power-law extrapolation below the regime where large sky
  surveys are currently complete.  The cyan band shows how the extrapolation
  would change as the faint-end slope of the galaxy stellar mass function ($\alpha$) is varied
  over the same range illustrated by the shaded gray band in Figure \ref{fig:mf}.
  Note that the enumeration of $\mstar = 10^5 \msun$ galaxies could provide a
  strong discriminator on faint-end slope, as the $\pm 0.15$ range in $\alpha$ shown
maps to an order of magnitude difference in the halo mass associated with this galaxy
  stellar mass and a corresponding
  factor of $\sim 10$ shift in the galaxy/halo counts shown in Figure \ref{fig:massfunc}.
  }
\label{fig:AM}
\end{figure}

At the small masses that most concern this review, dark matter halo counts
follow $dn/dM \propto M^{\alpha}$ with a steep slope $\alpha_{dm} \simeq -1.9$
compared to the observed stellar mass function slope of $\alpha_g = -1.47$
\citep[][which is consistent with the updated GAMA results shown in Figure \ref{fig:mf}]{baldry2012}. Current surveys that cover enough sky to provide a global
field stellar mass function reach a completeness limit of
$\mstar \approx 10^{7.5} \msun$.  At this mass, galaxy counts are more than two
orders of magnitude below the naive baryonic mass function $f_b \mvir$.  The
shaded band illustrates how the stellar mass function would extrapolate 
to the faint regime  spanning a range of
faint-end slopes $\alpha$ that are marginally consistent with
observations at the completeness limit.

One clear implication of this comparison is that galaxy formation efficiency
($\epsilon_\star$) must vary in a non-linear way as a function of $\mvir$ (at
least if \lcdm\ is the correct underlying model). Perhaps the cleanest way to
illustrate this is adopt the simple assumption of Abundance Matching (AM): that
galaxies and dark matter halos are related in a one-to-one way, with the most
massive galaxies inhabiting the most massive dark matter halos
\citep[][]{frenk1988,kravtsov2004a,conroy2006,moster2010,behroozi2013}.  The
results of such an exercise are presented in Figure \ref{fig:AM} (as derived by
Behroozi et al., in preparation).  The gray band shows the median
$\mstar - \mvir$ relation with an assumed 0.2 dex of scatter in $\mstar$ at
fixed $\mvir$.  The magenta band expands the scatter to 0.5 dex .  This relation
is truncated near the completeness limit in \citet{baldry2012}. The central
dashed line in Figure \ref{fig:AM} shows the median relation that comes from
extrapolating the \citet{baldry2012} mass function with their best-fit
$\alpha_g = -1.47$ down to the stellar mass regime of Local Group dwarfs.  The
cyan band brackets the range for the two other faint-end slopes shown in Figure
\ref{fig:mf}: $\alpha_g = -1.62$ and $-1.32$.

Figure \ref{fig:AM} allows us to read off the virial mass expectations for
galaxies of various sizes.  We see that Bright Dwarfs at the limit of detection
in large sky surveys ($\mstar \approx 10^8 \msun$) are naively associated with
$\mvir \approx 10^{11} \msun$ halos.  Galaxies with stellar masses similar to
the Classical Dwarfs at $\mstar \approx 10^6 \msun$ are associated with
$\mvir \approx 10^{10} \msun$ halos.  As we will discuss in Section
\ref{sec:solutions}, galaxies at this scale with $\mstar/\mvir \approx 10^{-4}$
are at the critical scale where feedback from star formation may not be
energetic enough to alter halo density profiles significantly.  Finally,
Ultra-faint Dwarfs with $\mstar\approx 10^4 \msun$, $\mvir\approx 10^{9} \msun$,
and $\mstar/\mvir \approx 10^{-5}$ likely sit at the low-mass extreme of galaxy
formation.

\begin{marginnote}[]
  \entry{Bright Dwarfs}{$\mstar\approx 10^{8} \msun$
    $\mvir \approx 10^{11}\msun$ $\mstar/\mvir \approx 10^{-3}$}
  \entry{Classical Dwarfs}{$\mstar\approx 10^6 \msun$
    $\mvir\approx 10^{10} \msun$ $\mstar/\mvir \approx 10^{-4}$}
  \entry{Ultra-faint Dwarfs}{$\mstar\approx 10^4 \msun$
    $\mvir\approx 10^{9} \msun$ $\mstar/\mvir \approx 10^{-5}$}
\end{marginnote}

\subsubsection{Kinematic Measures} An alternative way to connect to the dark
matter halo hosting a galaxy is to determine the galaxy's dark matter mass
kinematically.  This, of course, can only be done within a central radius probed
by the baryons.  For the small galaxies of concern for this review, extended
mass measurements via weak lensing or hot gas emission is infeasible.  Instead,
masses (or mass profiles) must be inferred within some inner radius, defined
either by the stellar extent of the system for dSphs and/or the outer rotation
curves for rotationally-supported gas disks.

Bright dwarfs, especially those in the field, often have gas disks with ordered
kinematics.  If the gas extends far enough out, rotation curves can be
extracted that extend as far as the flat part of the galaxy rotation curve
$V_{\rm flat}$.  If care is taken to account for non-trivial velocity
dispersions in the mass extraction \citep[e.g.,][]{kuzio-de-naray2008}, then we
can associate $V_{\rm flat} \approx \vmax$.

Owing to the difficulty in detecting them, the faintest galaxies known are all
satellites of the Milky Way or M31 and are dSphs.  These lack rotating gas
components, so rotation curve measurements are impossible.  Instead, dSphs are
primarily stellar dispersion-supported systems, with masses that are best probed
by velocity dispersion measurements obtained star-by-star for the closest dwarfs
\citep[e.g.,][]{walker2009,simon2011,kirby2014}.  For systems of this kind, the
mass can be measured accurately within the stellar half-light radius
\citep{walker2009}.  The mass within the de-projected (3D) half-light radius
($r_{1/2}$) is relatively robust to uncertainties in the stellar velocity
anisotropy and is given by $M(<r_{1/2}) = 3\, \sigma_\star^2 \, r_{1/2} / G$, where
$\sigma_\star$ is the  measured, luminosity-weighted line-of-sight velocity dispersion
\citep{wolf2010}.  This formula is equivalent to saying that the circular
velocity at the half-light radius is
$V_{1/2} = V_c(r_{1/2} )= \sqrt{3} \,\sigma_\star$.  The value of $V_{1/2}$
($\le \vmax$) provides a one-point measurement of the host halo's rotation curve
at $r=r_{1/2}$.

\subsection{Connections to particle physics}
\label{subsec:particle_physics}
Although the idea of ``dark" matter had been around since at least \citet{zwicky1933}, it was not until rotation curve measurements of galaxies in the 1970s revealed the need for significant amounts of non-luminous matter \citep{freeman1970, rubin1978, bosma1978, rubin1980} that dark matter was taken seriously by the broader astronomical community (and shortly thereafter, it was recognized that dwarf galaxies might serve as sensitive probes of dark matter; \citealt{aaronson1983, faber1983, lin1983}). Very quickly, particle physicists realized the potential implications for their discipline as well. Dark matter candidates were grouped into categories based on their effects on structure formation. ``Hot" dark matter (HDM) particles remain relativistic until relatively late in the Universe's evolution and smooth out perturbations even on super-galactic scales; ``warm" dark matter (WDM) particles have smaller initial velocities, become non-relativistic earlier, and suppress perturbations on galactic scales (and smaller); and CDM has negligible thermal velocity and does not suppress structure formation on any scale relevant for galaxy formation. Standard Model neutrinos were initially an attractive (hot) dark matter candidate; by the mid-1980s, however, this possibility had been excluded on the basis of general phase-space arguments \citep{tremaine1979}, the large-scale distribution of galaxies \citep{white1983a}, and properties of dwarf galaxies \citep{lin1983}. The lack of a suitable Standard Model candidate for particle dark matter has led to significant work on particle physics extensions of the Standard Model. From a cosmology and galaxy formation perspective, the unknown particle nature of dark matter means that cosmologists must make assumptions about dark matter's origins and particle physics properties and then
investigate the resulting cosmological implications. 
\begin{marginnote}[]
\entry{Cold Dark Matter (CDM)}{\\$m \sim 100 \,{\rm GeV}$, $v_{\rm th}^{z=0}
  \approx 0\,\kms$}
\entry{Warm Dark Matter (WDM)}{\\$m \sim 1 \,{\rm keV}$, $v_{\rm th}^{z=0} \sim 0.03 \,\kms$}
\entry{Hot Dark Matter (HDM)}{\\$m \sim 1\,{\rm eV}$, $v_{\rm th}^{z=0} \,\sim 30 \,\kms$}
\end{marginnote}

A general class of models
that are appealing in their simplicity is that of \textit{thermal relics}.
Production and destruction of dark matter particles are in equilibrium so long
as the temperature of the Universe $kT$ is larger than the mass of the dark
matter particle $m_{\rm DM}c^2$. At lower temperatures, the abundance is
exponentially suppressed, as destruction (via annihilation) dominates over
production. At some point, the interaction rate of dark matter particles drops
below the Hubble rate, however, and the dark matter particles ``freeze out'' at
a fixed number density (see, e.g., \citealt{kolb1990}; this is also known as chemical decoupling). Amazingly,
if the annihilation cross section is typical of weak-scale physics, the
resulting freeze-out density of thermal relics with $m\sim 100\,{\rm GeV}$ is
approximately equal to the observed density of dark matter today (e.g., \citealt{jungman1996}). This subset of
thermal relics is referred to as \textit{weakly-interacting massive particles
  (WIMPs)}. The observation that new physics at the weak scale naturally leads
to the correct abundance of dark matter in the form of WIMPs 
is known as the ``WIMP miracle'' \citep{feng2008} and has been the
basic framework for dark matter over the past 30 years.

WIMPs are not the only viable dark matter candidate, however, and it is
important to note that the WIMP miracle could be a red herring. \textit{Axions},
which are particles invoked to explain the strong CP problem of quantum
chromodynamics (QCD), and right-handed neutrinos (often called \textit{sterile
  neutrinos}), which are a minimal extension to the Standard Model of particle
physics that can explain the observed baryon asymmetry and why neutrino masses
are so small compared to other fermions, are two other hypothetical particles
that may be dark matter (among a veritable zoo of additional possibilities; see \citealt{feng2010} for a recent review). While WIMPs, axions, and sterile neutrinos are capable of producing the observed abundance of dark matter in the present-day Universe, they can have very different effects on the mass spectrum of cosmological perturbations.

While the cosmological perturbation spectrum is initially set by physics in the very early universe (inflation in the standard scenario), the microphysics of
dark matter affects the evolution of those fluctuations at later times. In the standard WIMP
paradigm, the low-mass end of the CDM hierarchy is set by first collisional
damping (subsequent to chemical decoupling but prior to kinetic decoupling of
the WIMPs), followed by free-streaming (e.g., \citealt{hofmann2001, bertschinger2006}). For typical 100 GeV WIMP candidates,
these processes erase cosmological perturbations with $M \la 10^{-6}\,\msun$
(i.e., Earth mass; \citealt{green2004}). Free-streaming also sets the low-mass end of the mass
spectrum in models where sterile neutrinos decouple from the plasma while
relativistic. In this case, the free-streaming scale can be approximated by the
(comoving) size of the horizon when the sterile neutrinos become
non-relativistic. The comoving horizon size at $z = 10^7$, corresponding to
$m\approx 2.5 \,{\rm keV}$, is approximately 50 kpc, which is significantly
smaller than the scale derived above for $L^*$ galaxies. keV-scale sterile
neutrinos are therefore observationally-viable dark matter candidates (see
\citealt{adhikari2016} for a recent, comprehensive review). QCD axions are
typically $\sim \mu{\rm eV}$-scale particles but are produced out of thermal
equilibrium \citep{kawasaki2013}. Their free-streaming scale is significantly smaller than that of a typical WIMP (see Section~\ref{subsubsec:linear}).

The previous paragraphs have focused on the effects of collisionless damping and
free-streaming -- direct consequences of the particle nature of dark matter --
in the linear regime of structure formation. Dark matter microphysics can also
affect the non-linear regime of structure formation. In particular, dark matter
self-interactions -- scattering between two dark matter particles -- will affect
the phase space distribution of dark matter.  Within observational constraints,
dark matter self-interactions could be relevant in the dense centers of dark
matter halos. By transferring kinetic energy from high-velocity particles to
low-velocity particles, scattering transfers ``heat'' to the centers of dark
matter halos, reducing their central densities and making their velocity
distributions nearly isothermal. This would have a direct effect on galaxy
formation, as galaxies form within the centers of dark matter halos and the
motions of their stars and gas trace the central gravitational potential. These
effects are discussed further in Section~\ref{subsubsec:nonlinear}.

The particle nature of dark matter is therefore reflected in the cosmological
perturbation spectrum, in the abundance of collapsed dark matter structures as a
function of mass, and in the density and velocity distribution of dark matter in
virialized dark matter halos.

\begin{textbox}[ht]\section{THREE CHALLENGES TO BASIC \bmlcdm\ PREDICTIONS}
  There are three classic problems associated with the small-scale predictions
  for dark matter in the \lcdm\ framework.  Other anomalies exist, including
  some that we discuss in this review, but these three are important because 1)
  they concern basic predictions about dark matter that are fundamental to the
  hierarchical nature of the theory; and 2) they have received significant
  attention in the literature.

\subsection{Missing Satellites and Dwarfs [Figures~\ref{fig:massfunc}--\ref{fig:MSP_AM}]}
The observed stellar mass functions of field galaxies and satellite galaxies in
the Local Group is much flatter at low masses than predicted dark matter halo
mass functions: $dn/d\mstar \propto \mstar^{\alpha_g}$ with
$\alpha_{g} \simeq -1.5$ (vs. $\alpha \simeq -1.9$ for dark matter).  The issue
is most acute for Galactic satellites, where completeness issues are less of a
concern.  There are only $\sim 50$ known galaxies with $M_\star > 300 \msun$
within $300$ kpc of the Milky Way compared to as many as $\sim 1000$ dark
subhalos (with $M_{\rm sub} > 10^{7} \msun$) that could conceivably host
galaxies.  One solution to this problem is to posit that galaxy formation
becomes increasingly inefficient as the halo mass drops.  The smallest dark
matter halos have simply failed to form stars altogether.

\subsection{Low-density Cores vs.  High-density Cusps [Figure~\ref{fig:cuspcore}]}
The central regions of dark-matter dominated galaxies as inferred from rotation
curves tend to be both less dense (in normalization) and less cuspy (in inferred
density profile slope) than predicted for standard \lcdm\ halos (such as those
plotted in Figure~\ref{fig:nfw}).  An important question is whether baryonic feedback alters
the structure of dark matter halos.

\subsection{Too-Big-to-Fail [Figure~\ref{fig:tbtf}]} 
The local universe contains too few galaxies with central densities indicative
of $\mvir \simeq 10^{10} \msun$ halos.  Halos of this mass are generally
believed to be too massive to have failed to form stars, so the fact that they
are missing is hard to understand.  The stellar mass associated with this halo
mass scale ($M_\star \simeq 10^6 \msun$, Figure \ref{fig:AM}) may be too small
for baryonic processes to alter their halo structure (see Figure
\ref{fig:feedback}).
  
  \end{textbox}

\section{OVERVIEW OF PROBLEMS}
\label{sec:problems}
The CDM paradigm as summarized in the previous section emerged among other
dark matter variants in the early 1980s \citep{peebles1982,blumenthal1984,davis1985}
with model parameters gradually settling to their current precise state (including $\Lambda$) in the
wake of overwhelming evidence from large-scale galaxy clustering, supernovae
measurements of cosmic acceleration, and cosmic microwave background studies,
among other data. The 1990s saw the first $N$-body simulations to resolve the
internal structure of CDM halos on small scales.  Almost immediately researchers
pinpointed the two most well-known challenges to the theory: the cusp-core
problem \citep{flores1994,moore1994} and the missing satellites problem
\citep{klypin1999,moore1999}.  This section discusses these two classic issues
from a current perspective goes on to describe a third problem, \tbtf\ \citep{boylan-kolchin2011}, which is in some sense is a confluence of the
first two. Finally, we conclude this section with a more limited discussion of
two other challenges faced by \lcdm\ on small scales: the apparent planar
distributions seen for Local Group satellites and the dynamical scaling
relations seen in galaxy populations.

\subsection{Missing Satellites}
\label{subsec:msp}
\begin{figure}[t]
 \begin{minipage}{.50\textwidth}
\includegraphics[width=\linewidth]{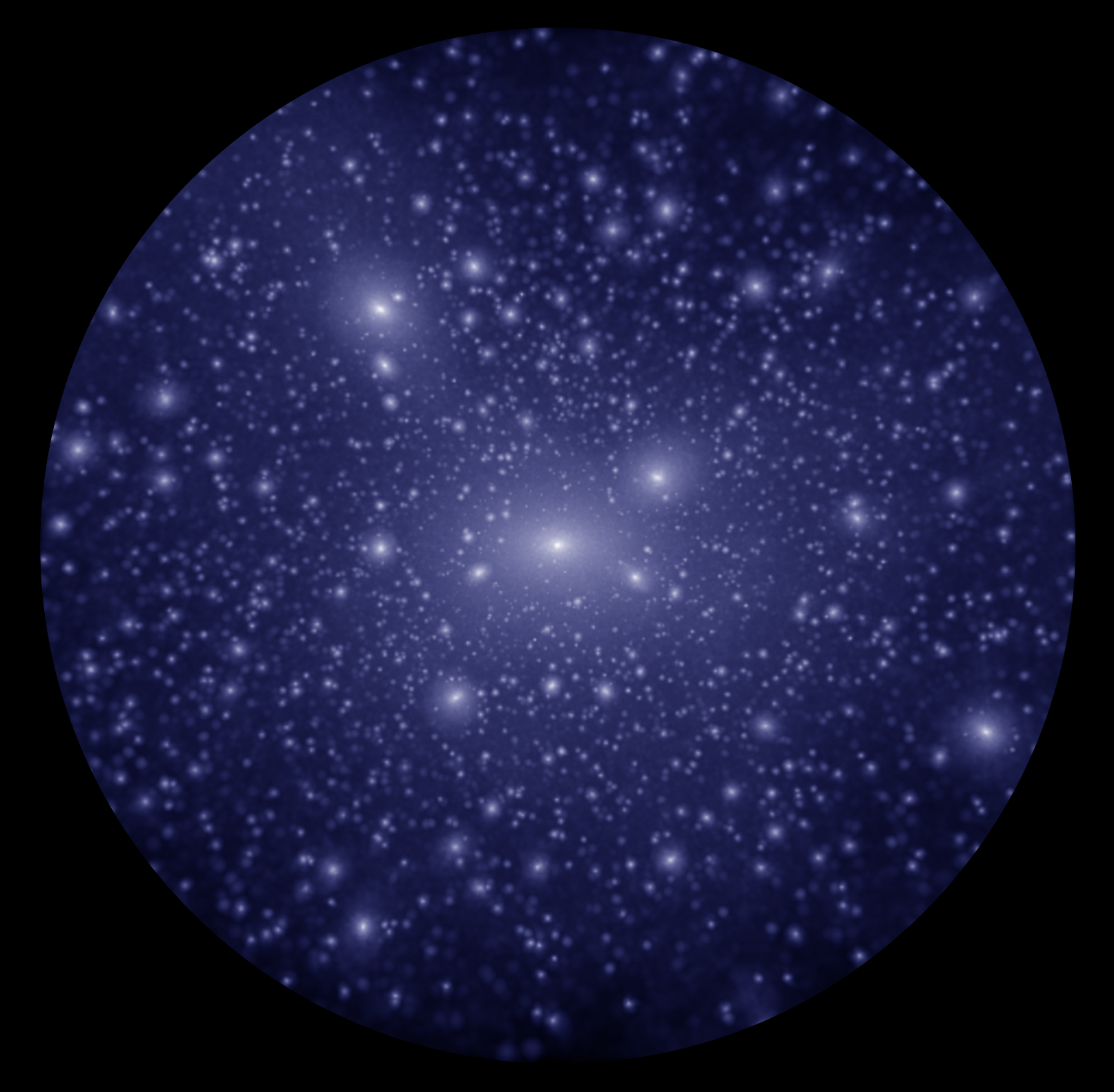}
 \end{minipage}
 \begin{minipage}{.49\textwidth}
\includegraphics[width=\linewidth]{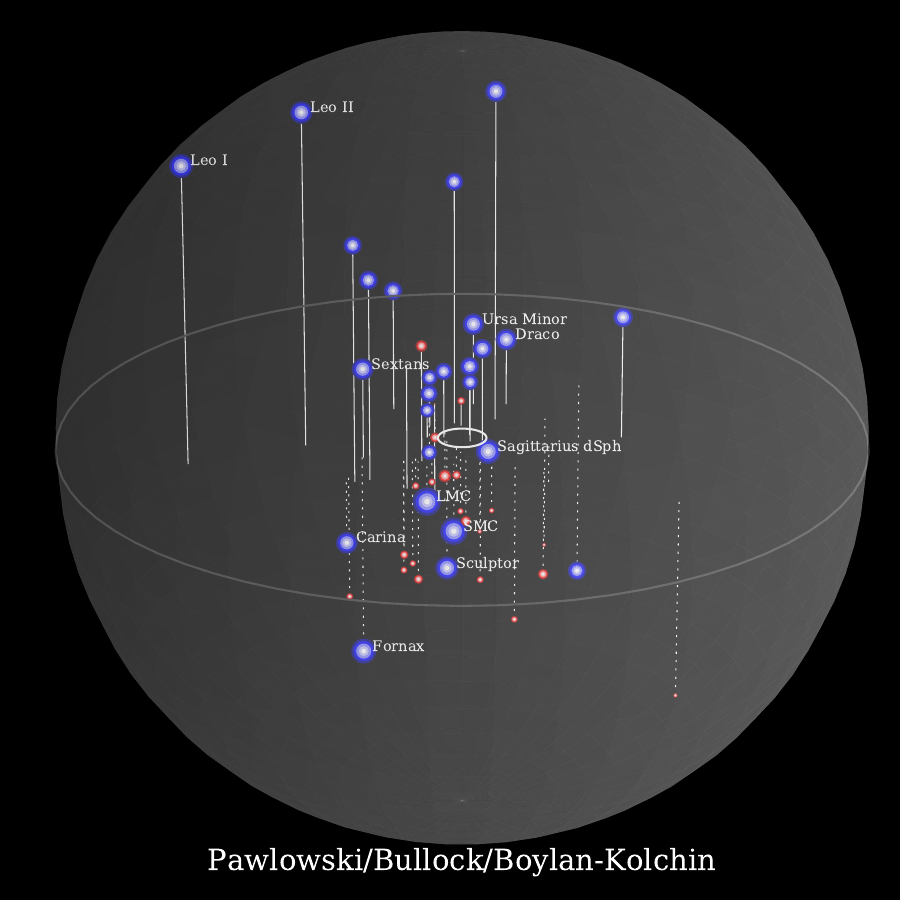}
 \end{minipage}
\caption{The Missing Satellites Problem:  Predicted \lcdm\ substructure (left) vs. known Milky Way satellites
  (right).   The image on the left shows the \lcdm\ dark matter distribution within a sphere of radius 250 kpc around the center of a Milky-Way size dark matter halo  (simulation by V. Robles and T. Kelley in collaboration with the authors). The image on the right (by M. Pawlowski in collaboration with the authors)
  shows the current census of Milky Way satellite galaxies, with galaxies discovered since 2015 in red. The Galactic disk is represented by a circle of radius 15 kpc at the center and the outer sphere has a radius of 250 kpc.  The 11 brightest (classical) Milky Way satellites are labeled by name.  Sizes of the symbols are not to scale but are rather proportional to the log of each satellite galaxy's stellar mass.  Currently, there are $\sim 50$ satellite galaxies of the Milky Way compared to thousands of predicted subhalos with $\mpeak \ga 10^7\,\msun$. }
\label{fig:satellites}
\end{figure}

The highest-resolution cosmological simulations of MW-size halos in the \lcdm\,
paradigm have demonstrated that dark matter (DM) clumps exist at all resolved
masses, with no break in the subhalo mass function down to the numerical
convergence limit
\citep[e.g.,][]{springel2008,VL2,GHALO,garrison-kimmel2014,Griffen2016}. We
expect thousands of subhalos with masses that are (in principle) large enough to
have supported molecular cooling ($\mpeak \gtrsim 10^7~\msun$).  Meanwhile,
only $\sim 50$ satellite galaxies down to $\sim 300~\msun$ in stars are known to orbit
within the virial radius of the Milky Way \citep{drlica-wagner2015}.  Even though there
is real hope that future surveys could bring the census of ultra-faint dwarf
galaxies into the hundreds \citep{tollerud2008, hargis2014}, it seems unlikely there are
thousands of undiscovered dwarf galaxies to this limit within the virial volume of the Milky
Way.  The current situation is depicted in Figure \ref{fig:satellites}, which
shows the dark matter distribution around a Milky Way size galaxy as predicted by a \lcdm\ simulation next to a map of the known galaxies of the Milky Way on the same scale.

Given the discussion of abundance matching in Section \ref{sec:AM} and the
associated Figure~\ref{fig:AM}, it is reasonable to expect that dark matter
halos become increasingly inefficient at making galaxies at low masses and at
some point go completely dark.  Physical mass scales of interest in this regard
include the mass below which
reionization UV feedback likely suppresses gas accretion
$\mvir \approx 10^9 \, \msun$ \citep[$\vmax \gtrsim 30 \, \kms$; e.g.,][]{efstathiou1992,bullock2000, benson2002, bovill2009, sawala2016a} and the minimum mass for atomic cooling in the early Universe, $\mvir \approx 10^8 \, \msun$
\citep[$\vmax \gtrsim 15 \, \kms$; see, e.g.,][]{rees1977}.
According to Figure \ref{fig:AM}, these physical effects are likely to become 
dominant in the regime of ultra-faint galaxies $\mstar \lesssim 10^5 \msun$.

\begin{figure}[t]
  \includegraphics[width=0.7\textwidth]{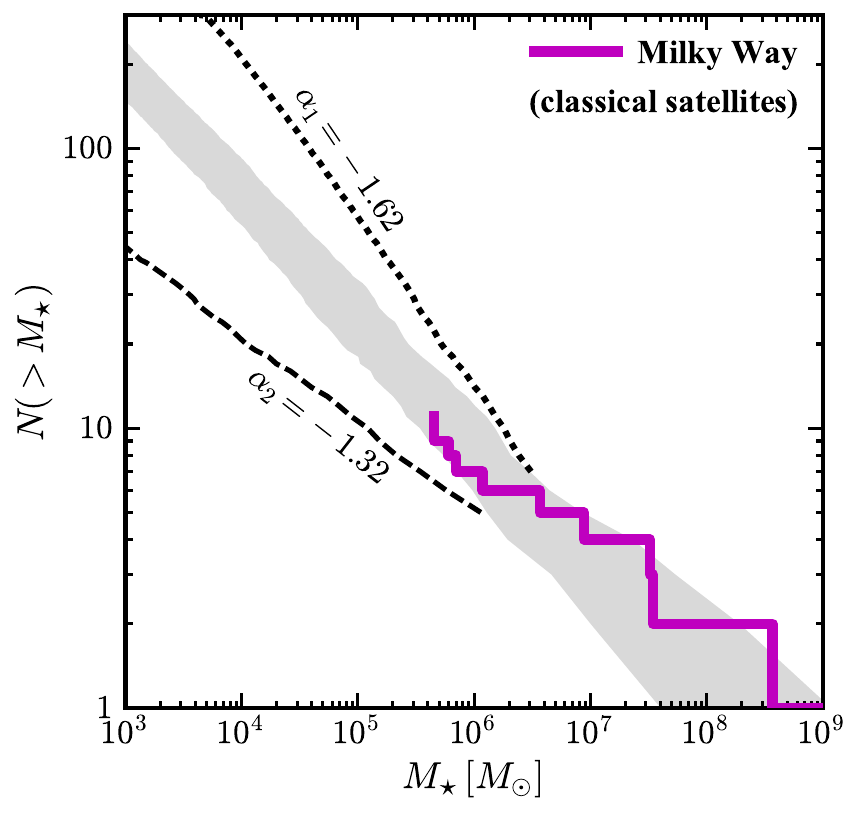}
  \caption{``Solving" the Missing Satellites Problem with abundance matching. The cumulative count of dwarf galaxies around the Milky Way (magenta) plotted down to completeness limits from
    \citet{garrison-kimmel2017}.  The gray shaded region shows the predicted
    stellar mass function from the dark-matter-only ELVIS simulations
    \citep{garrison-kimmel2014} combined with the fiducial AM relation shown in
    Figure \ref{fig:AM}, assuming zero scatter. If the faint end slope of the stellar mass function is shallower (dashed) or steeper (dotted), the predicted abundance of satellites with $\mstar > 10^4\,\msun$ throughout the Milky Way's virial volume differs by a factor of 10. Local Group counts can therefore serve as strong constraints on galaxy formation models.}
\label{fig:MSP_AM}
\end{figure}

The question then becomes: can we simply adopt the abundance-matching relation derived from
field galaxies to ``solve" the Missing Satellites Problem down to the scale of
the classical MW satellites (i.e., $\mvir \simeq 10^{10} \msun$ $\leftrightarrow$ $\mstar \simeq 10^6 \msun$)?  Figure \ref{fig:MSP_AM} \citep[modified from][]{garrison-kimmel2017} shows that the answer is
likely ``yes."  Shown in magenta is the cumulative count of Milky Way
satellite galaxies  within 300 kpc of the Galaxy plotted down to the stellar mass completeness limit within that volume.  The shaded band shows the $68\%$ range
predicted stellar mass functions from the dark-matter-only ELVIS simulations \citep{garrison-kimmel2014} combined with the AM relation shown in Figure \ref{fig:AM} with zero scatter.  The agreement is not perfect, but there is no over-prediction.
The dashed lines show  how the predicted satellite stellar mass  functions would 
change for different assumed (field galaxy) faint-end slopes in the calculating the AM relation. An important avenue going forward will be to push these comparisons down to
the ultra-faint regime, where strong baryonic feedback effects are expected to
begin shutting down galaxy formation altogether.

\begin{figure}[!htb]
\includegraphics[width=0.8\textwidth]{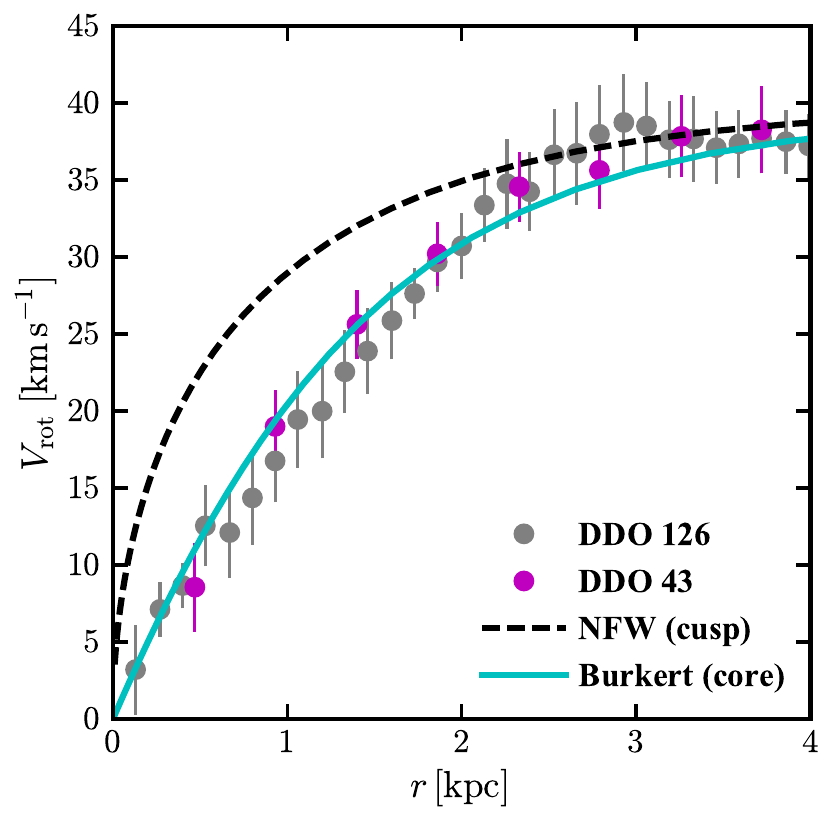}
\caption{The Cusp-Core problem. The dashed line shows the naive \lcdm\ expectation (NFW, from dark-matter-only simulations) for a typical rotation curve of a $\vmax \approx 40 \, \kms$ galaxy.  This rotation curve
rises quickly, reflecting a central density profile that rises as a cusp with $\rho \propto 1/r$.  The data points show  
the rotation curves of two example galaxies of this size from the LITTLE THINGS survey \citep{oh2015}), which are more slowly rising and better fit by a density profile with a constant density core \citep[][cyan line]{Burkert1995}.}
\label{fig:cuspcore}
\end{figure}

\subsection{Cusp, Cores, and Excess Mass}
 
As discussed in Section 1, \lcdm\ simulations that include only dark matter 
predict that dark matter halos should have density profiles that rise steeply at
small radius $\rho(r) \propto r^{-\gamma}$, with $\gamma \simeq 0.8-1.4$ over
the radii of interest for small galaxies \citep{navarro2010}.  This is in
contrast to many (though not all) low-mass dark-matter-dominated galaxies with
well-measured rotation curves, which prefer fits with constant-density cores
($\gamma \approx 0-0.5$; e.g., 
\citealt{McGaugh2001,Marchesini2002,simon2005,deBlok2008,kuzio-de-naray2008}). A
related issue is that fiducial \lcdm\ simulations predict more dark matter in
the central regions of galaxies than is measured for the galaxies that they
should host according to AM.  This ``central density problem" is an issue of
normalization and exists independent of the precise slope of the central density
profile \citep{alam2002,oman2015}.  While these problems are in principle
distinct issues, as the second refers to a tension in total cumulative mass and
the first is an issue with the derivative, it is likely that they point to a common
tension.  Dark-matter-only \lcdm\ halos are too dense and too cuspy in their
centers compared to many observed galaxies.

Figure \ref{fig:cuspcore} summarizes the basic problem.  Shown as a dashed line
is the typical circular velocity curve predicted for an NFW \lcdm~ dark matter
halo with $\vmax \approx 40 \kms$ compared to the observed 
rotation curves for two galaxies with the same asymptotic velocity
from \citet{oh2015}.  The observed rotation curves rise
much more slowly than the \lcdm\ expectation, reflecting core densities that are
lower and more core-like than the fiducial prediction.

\subsection{Too-Big-To-Fail}
\begin{figure}[!htb]
\begin{minipage}{0.5\textwidth}
\includegraphics[width=0.99\linewidth]{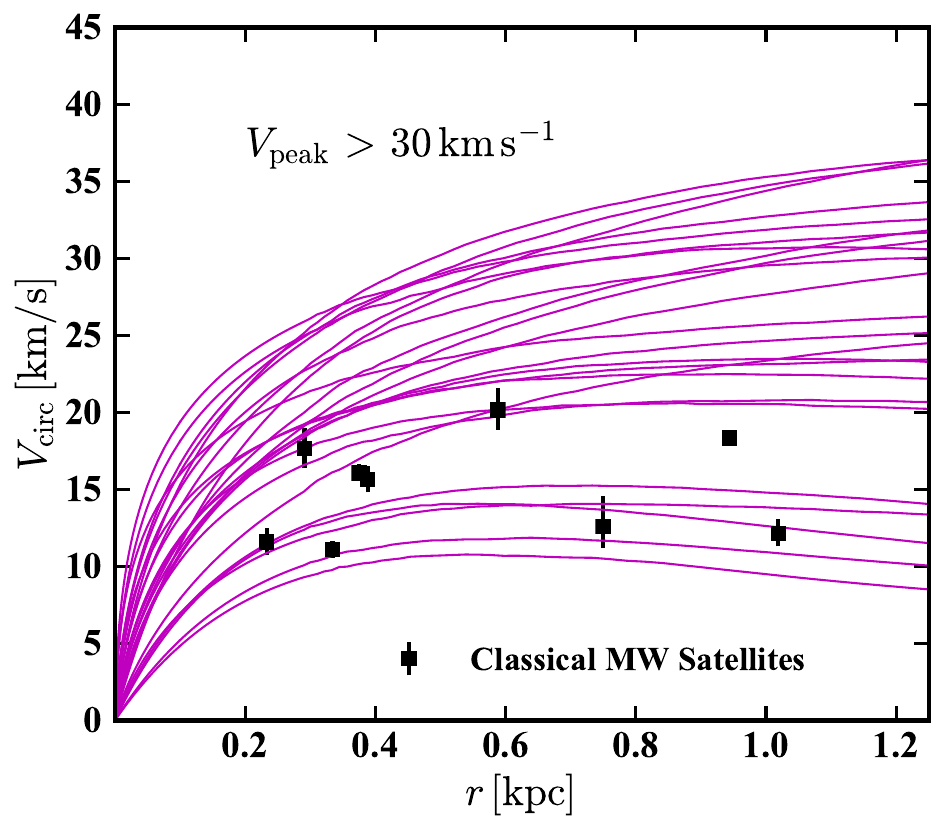}
\end{minipage}
\begin{minipage}{0.5\textwidth}
\includegraphics[width=0.99\linewidth]{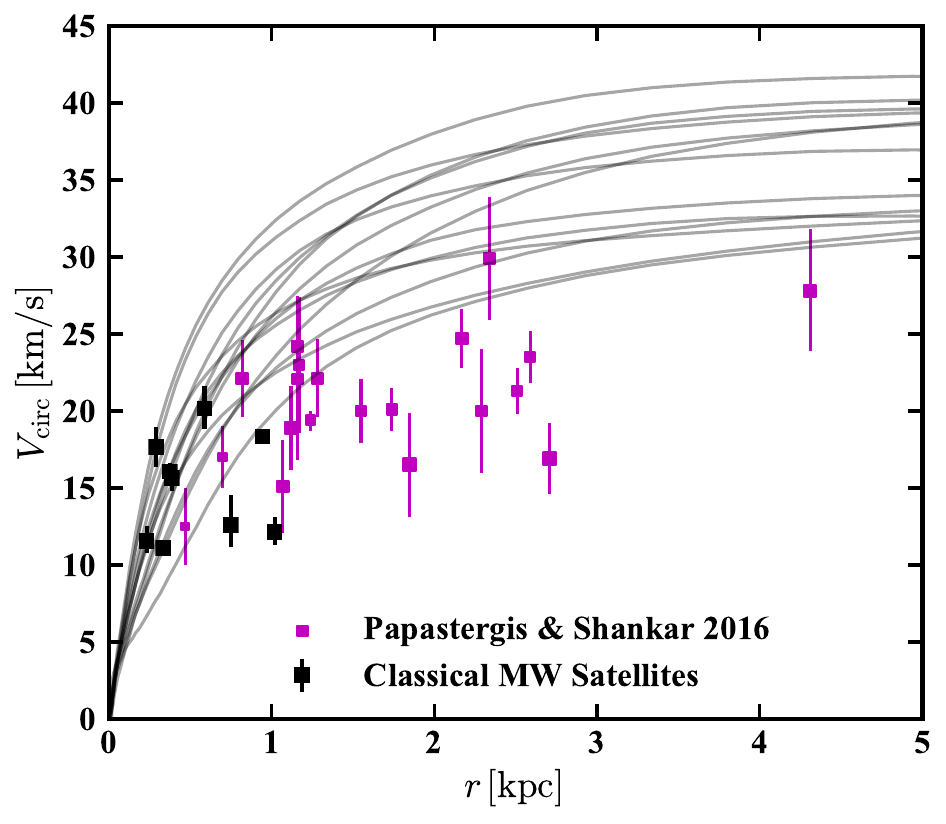}
\end{minipage}
\caption{The Too-Big-to-Fail Problem. {\em Left:} Data points show the circular
  velocities of classical Milky Way satellite galaxies with
  $\mstar \simeq 10^{5-7} \msun$ measured at their half-light radii
  $r_{1/2}$. The magenta lines show the circular velocity curves of subhalos
  from one of the (dark matter only) Aquarius simulations.  These are
  specifically the subhalos of a Milky Way-size host that have peak maximum
  circular velocities $\vmax > 30 \, \kms$ at some point in their histories.
  Halos that are this massive are likely resistant to strong star formation
  suppression by reionization
  and thus naively too big to have failed to form stars \citep[modified
  from][]{boylan-kolchin2012}.  The existence of a large population of such
  satellites with greater central masses than any of the Milky Way's dwarf
  spheroidals is the original Too-Big-to-Fail problem. {\it Right:} The same
  problem -- a mismatch between central masses of simulated dark matter systems
  and observed galaxies -- persists for field dwarfs (magenta points), indicating it is not a
  satellite-specific process (modified from \citealt{papastergis2017}). The field 
  galaxies shown all have stellar masses in the range $5.75 \leq \log_{10}(\mstar/\msun) \leq 7.5$.  The gray curves
  are predictions for \lcdm\ halos from the fully self-consistent hydrodynamic
  simulations of \citet{fitts2016} that span the same stellar mass range in the simulations
  as the observed galaxies.}
  \label{fig:tbtf}
\end{figure}

\label{subsec:tbtf}
As discussed above, a straightforward and natural solution to the missing
satellites problem within \lcdm\ is to assign the known Milky Way satellites to
the largest dark matter subhalos (where largest is in terms of either
present-day mass or peak mass) and attribute the lack of observed galaxies in
in the remaining smaller subhalos to galaxy formation physics. As pointed out by
\citet{boylan-kolchin2011}, this solution makes a testable prediction: the
inferred central masses of Milky Way satellites should be consistent with the
central masses of the most massive subhalos in \lcdm\ simulations of Milky
Way-mass halos. Their comparison of observed central masses to \lcdm\
predictions from the
Aquarius \citep{springel2008} and Via Lactea II \citep{diemand2008} simulations revealed
that 
the most massive \lcdm\ subhalos were systematically too centrally dense to host the
bright Milky Way satellites \citep{boylan-kolchin2011,
  boylan-kolchin2012}. While there are subhalos with central masses
comparable to the Milky Way satellites,
these subhalos were never among the $\sim10$ most massive
(Figure~\ref{fig:tbtf}). Why would galaxies fail to form in the most massive
subhalos, yet form in dark matter satellites of lower mass? The most massive
satellites should be ``too big to fail'' at forming galaxies if the lower-mass
satellites are capable of doing so (thus the origin of the name of this
problem).  
In short, while the \textit{number} of massive subhalos in dark-matter-only simulations matches the number of classical dwarfs observed (see Figure 8), the \textit{central densities} of these simulated dwarfs are higher than the central densities observed in the real galaxies (see Figure 10).  

While too-big-to-fail was originally identified for satellites of the Milky
Way, it was subsequently found to exist in Andromeda \citep{tollerud2014} and
field galaxies in the Local Group (those outside the virial radius of the Milky
Way and M31; \citealt{kirby2014}). Similar discrepancies were also pointed out
for more isolated low-mass galaxies, first based on HI rotation curve data
\citep{ferrero2012} and subsequently using velocity width measurements
\citep{papastergis2015, papastergis2016}. This version of \tbtf\ in the field is also manifested in the velocity function of field galaxies\footnote{We note that the mismatch between the observed and predicted velocity function can also be interpreted as a ``missing dwarfs" problem if one considers the discrepancy as one in numbers at fixed $V_{\rm halo}$. We believe, however, that the more more plausible interpretation is a discrepancy in $V_{\rm halo}$ at fixed number density.} (\citealt{zavala2009,klypin2015, trujillo-gomez2016,schneider2016}, though see \citealt{maccio2016} and \citealt{brooks2017} for arguments that no discrepancy exists). The generic observation in the
low-redshift Universe, then, is that the inferred central masses of galaxies
with $10^5 \la \mstar/\msun \la 10^8$ are $\sim 50\%$ smaller than expected from
dissipationless \lcdm\ simulations.

The too-big-to-fail and core/cusp problems would be naturally connected if
low-mass galaxies generically have dark matter cores, as this would reduce their
central densities relative to CDM expectations\footnote{For a sense of the problem, the amount of mass that
would need to be removed to alleviate the issue on classical dwarf scales is $\sim 10^7 \msun$ within  $\sim 300$ pc}. 
However, the problems are, in
principle, separate: one could imagine galaxies that have large constant-density cores yet still with 
too much central mass relative to CDM predictions (solving the
core/cusp problem but not too-big-to-fail), or having cuspy profiles with
overall lower density amplitudes than CDM (solving too-big-to-fail but not
core/cusp).

\begin{figure}[t]
 \begin{minipage}{.5\textwidth}
\includegraphics[scale=0.4]{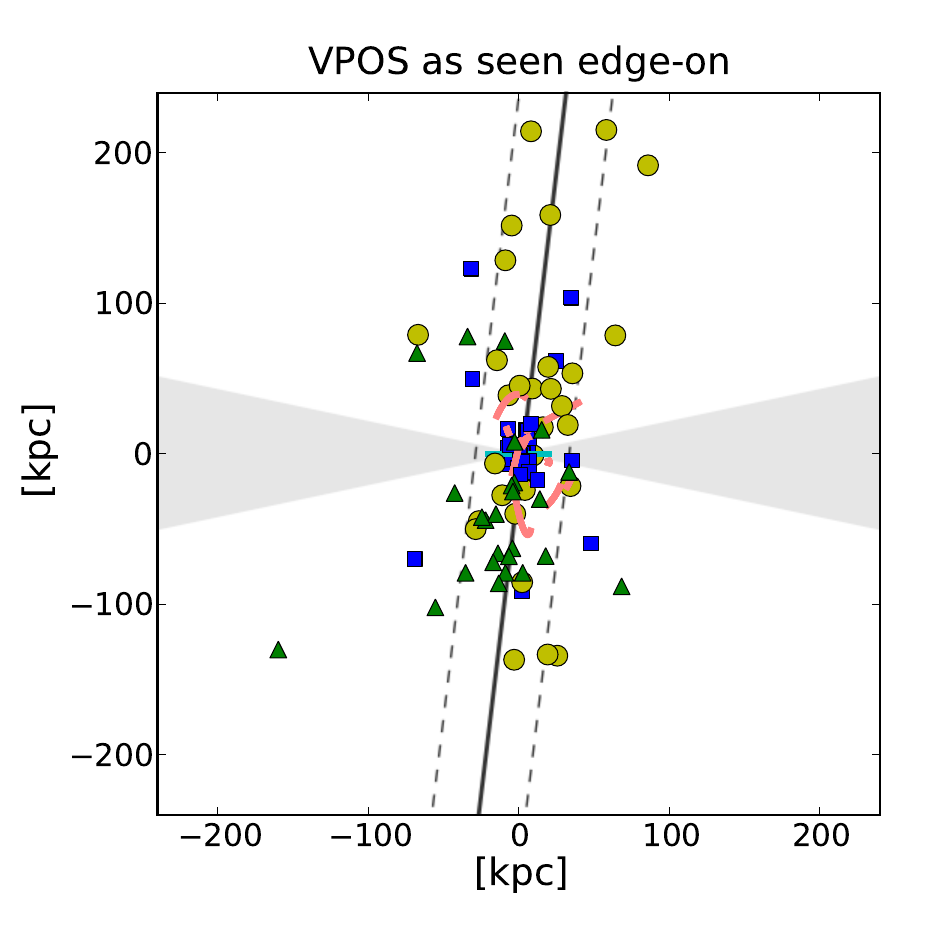}
 \end{minipage}
 \begin{minipage}{.5\textwidth}
\includegraphics[scale=0.4]{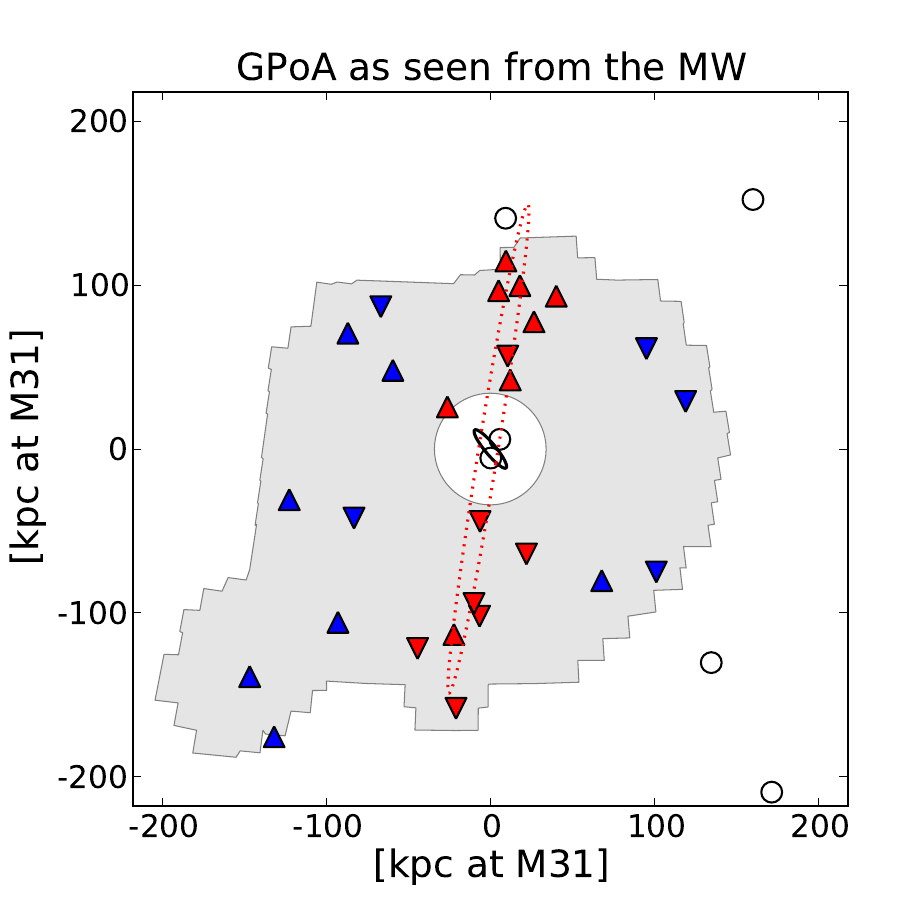}
 \end{minipage}
\caption{Planes of Satellites. \textit{Left}: Edge-on view of the satellite distribution around the Milky Way
  \citep[updated from][]{pawlowski2015} with the satellite galaxies in yellow,  young halo globular clusters and star clusters in blue, and all other newly-discovered objects (unconfirmed dwarf galaxies or star clusters) are shown as green triangles. The red lines in the center dictate the position and orientation of streams in the MW halo. The gray wedges span 24 degrees about the plane of the MW disk, where satellite discovery might be obscured by the Galaxy.
  \textit{Right}: The satellite distribution around Andromeda \citep[modified by M. Pawlowski from][]{ibata2013} where the red points
  are satellites belonging to the identified kinematic plane. Triangles pointing up are receding relative to M31.  Triangles pointing down are approaching. 
  }
\label{fig:planes}
\end{figure}

\subsection{Satellite Planes}
\label{subsec:planes}
\citet{kunkel1976} and \citet{lynden-bell1976} pointed out that satellite galaxies 
appeared to lie in a polar great circle around the Milky Way.
Insofar as this cannot be explained in a theory of structure formation, this observation
pre-dates all other small-scale structure issues in the Local Group by
approximately two decades. The anisotropic distribution of Galactic satellites
received scant attention until a decade ago, when \citet{kroupa2005} argued that
it proved that satellite galaxies cannot be related to dark matter substructures
(and thereby constituted another crisis for CDM). Kroupa et al.~examined classical,
pre-SDSS dwarf galaxies in and around the Milky Way and found that the observed
distribution was strongly non-spherical.  From this analysis, based
on the distribution of angles between the normal of the best-fitting plane of
dwarfs and the position vector of each MW satellite in the Galacto-centric
reference frame, Kroupa et al.~argued that 
  ``the mismatch between the number and spatial distribution of MW dwarves
  compared to the theoretical distribution challenges the claim that the MW
  dwarves are cosmological sub-structures that ought to populate the MW halo.''

This claim was quickly disputed by \citet{zentner2005}, who investigated the
spatial distribution of dark matter subhalos in simulated CDM halos and determined
that it was highly inconsistent with a spherical distribution. They found that
the planar distribution of MW satellites was marginally consistent with being a
random sample of the subhalo distributions in their simulations, and furthermore,
the distribution of satellites they considered likely to be luminous
(corresponding to the more massive subhalos) was even more consistent with
observations. A similar result was obtained at roughly the same time by
\citet{kang2005}. Slightly later, \citet{metz2007} argued that the distribution of MW
satellite galaxies was inconsistent, at the 99.5\% level, with isotropic or
prolate substructure distributions (as might be expected in \lcdm). 

Related analysis of  Milky Way satellite objects has further supported the idea that the configuration is highly unusual compared to \lcdm\ subhalo distributions \citep{pawlowski2012}, with the 3D motions of satellites suggesting that there is a preferred orbital pole aligned perpendicular to the observed spatial plane \citep{pawlowski2013a}. The left hand side of Figure \ref{fig:planes} shows the current distribution of satellites (galaxies and star clusters) around the Milky Way looking edge-on at the planar configuration.  Note that the disk of the Milky Way could, in principle, bias discoveries away from the MW disk axis, but it is not obvious that the orbital poles would be biased by this effect.  Taken together, the orbital poles and spatial configuration of MW satellites is highly unusual for a randomly drawn sample of \lcdm\ subhalos \citep{Pawlowski2015a}.

As shown in the right-hand panel of Figure 11, the M31 satellite galaxies also show evidence for having a
disk-like configuration \citep{metz2007}. 
Following the discovery of new M31 satellites
and the characterization of their velocities, \citet{conn2013} and
\citet{ibata2013} presented evidence that 15 of 27 Andromeda dwarf galaxies
indeed lie in a thin plane, and further, that that the southern satellites are mostly approaching us with respect to M31, while the northern satellites are mostly receding (as coded by the direction of the red triangles in Figure 11). This
suggests that the plane could be rotationally
supported. Our view of this plane is essentially edge-on, meaning we have
excellent knowledge of in-plane motions and essentially no knowledge of
velocities perpendicular to the plane. Nevertheless, even a transient plane of this kind would be exceedingly rare for \lcdm\ subhalos \citep[e.g.,][]{ahmed2017}.

Work in a similar vein has argued for the existence of planar
structures in the Centaurus A group \citep{tully2015} and for
rotationally-supported systems of satellites in a statistical sample of galaxies
from the SDSS \citep{ibata2015}. \citet{Libeskind2015} have used \lcdm\ simulations to suggest that some alignment of satellite systems in the local Universe may be naturally explained by the ambient shear field, though they cannot explain thin planes this way.  Importantly, \citet{Phillips2015} have re-analyzed the SDSS data and argued that it is not consistent with a ubiquitous co-rotating satellite population and rather more likely a statistical fluctuation.   More data that enables a statistical sample of hosts down to fainter satellites will be needed to  determine whether the configurations seen in the Local Group are common.

\begin{figure}[t!]
 \begin{minipage}{.5\textwidth}
\includegraphics[width=0.95\linewidth]{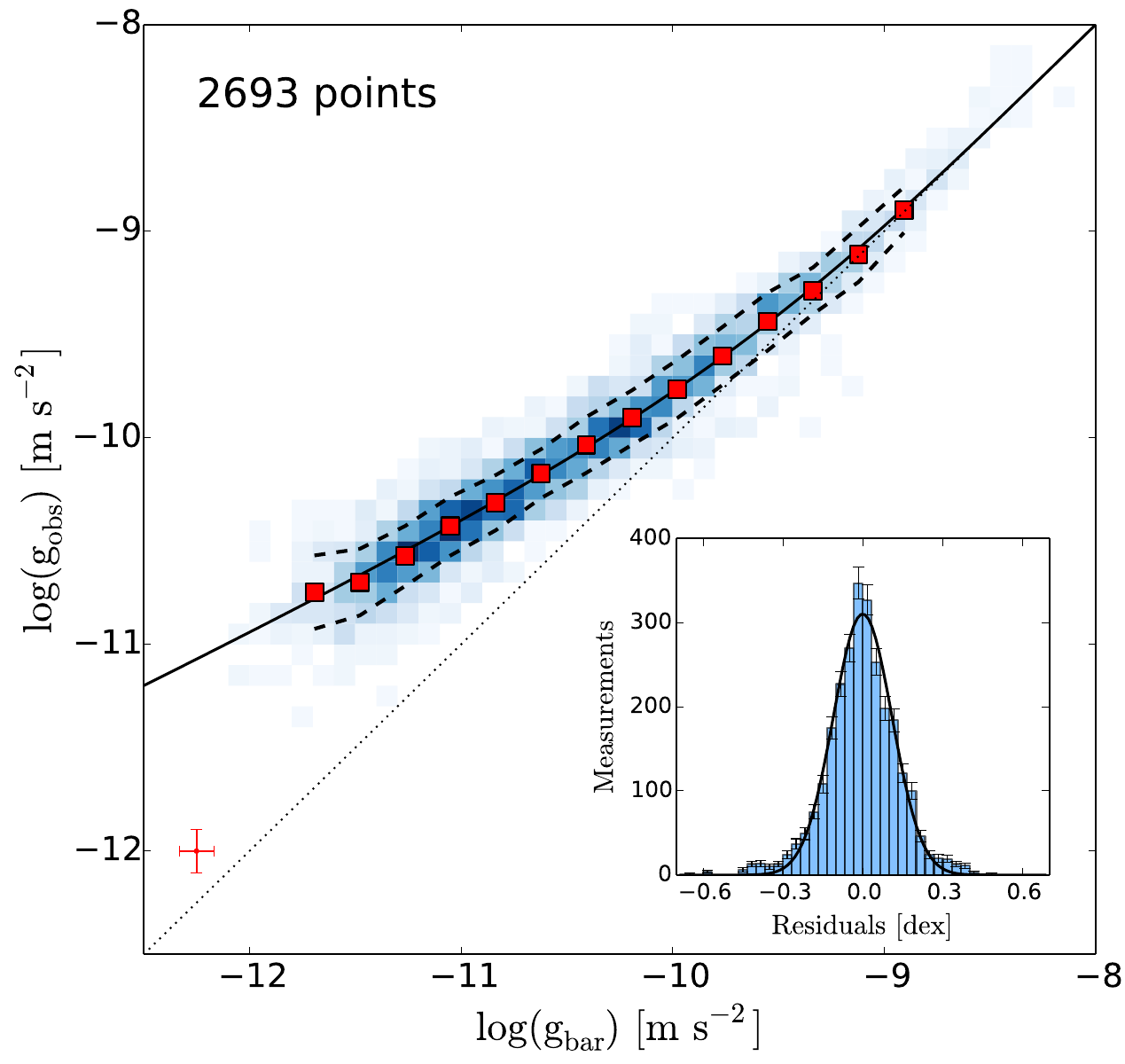}
 \end{minipage}
 \begin{minipage}{.5\textwidth}
\includegraphics[width=0.95\linewidth]{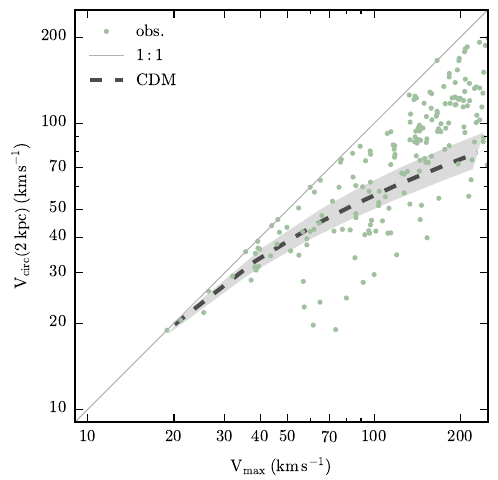}
 \end{minipage}
\caption{Regularity vs. Diversity. {\em Left:} The radial acceleration relation from \citet[][slightly modified]{McGaugh2016} showing the centripetal acceleration observed in rotation curves, $g_{\rm obs} =  V^2/r$, plotted versus the expected acceleration from observed baryons $g_{\rm bar}$ for 2700 individual data points from 153 galaxy rotation curves.   Large squares show the mean and the dashed line lines show the rms width.   {\em Right:}  Green points show the circular velocities of observed
  galaxies measured at $2~\kpc$ as a function of $\vmax$
  from \citet{oman2015} as re-created by \citet{creasey2017}.  For comparison, the gray band shows expectations from dark
  matter only \lcdm\ simulations. There is much more scatter at fixed $\vmax$ than
  predicted by the simulations.  Note that the galaxies used in the RAR in left-hand panel have $\vmax$ values that span the range shown on the right. The
  tightness of the acceleration relation is remarkable (consistent with zero scatter given observational error, red cross), especially given the
  variation in central densities seen on the right.}
\label{fig:scalings}
\end{figure}

\subsection{Regularity in the Face of Diversity}
Among the more puzzling aspects of galaxy phenomenology in the context of \lcdm\ are
the tight scaling relations between dynamical properties and baryonic
properties, even within systems that are dark matter dominated.  One well-known
example of this is the baryonic Tully-Fisher relation \citep{mcgaugh2012}, which
shows a remarkably tight connection between the total baryonic mass of a galaxy (gas plus
stars) and its circular velocity $V_{\rm flat}$ ($\simeq \vmax$): $M_b \propto V_{\rm flat}^4$.
Understanding this correlation with \lcdm\ models requires care for the low-mass
galaxies of most concern in this review \citep{brook2016}.

A generalization of the baryonic Tully-Fisher relation known as the radial acceleration relation (RAR) was recently introduced by 
\citet{McGaugh2016}.  Plotted in left-hand Figure \ref{fig:scalings}, the RAR
shows a tight correlation between the radial acceleration traced by rotation curves ($g_{\rm obs} = V^2/r$) and that predicted solely by the observed distribution of baryons ($g_{\rm bar}$)\footnote{This type of relation is what is generally expected in MOND, though the precise shape of the relation depends on the MOND interpolation function assumed (see \citealt{McGaugh2016} for a brief discussion).}. The upper right ``high-acceleration" portion of the relation correspond to baryon-dominated regions of (mostly large) galaxies.  Here the relation tracks the one-to-one line, as it must.  However, rotation curve points begin to peel away from the line, towards an acceleration larger than what can be explained by the baryons alone below a characteristic acceleration of $a_0 \simeq 10^{-10}$ m s$^{-2}$.  It is this additional acceleration that we attribute to dark matter.  The outer parts of large galaxies contribute to this region, as do virtually all parts of small galaxies.  It is surprising, however, that the dark matter contribution in the low-acceleration regime tracks the baryonic distribution so closely, particularly in light of the diversity in galaxy rotation curves seen among galaxies of at a fixed $V_{\rm flat}$, as we now discuss. 

The right-hand panel of Figure \ref{fig:scalings} illustrates the diversity in rotation curve shapes seen from galaxy to galaxy. Shown is a slightly modified version of a figure introduced by \citet{oman2015} and recreated by Creasey et al. (2017).  Each data point
corresponds to a single galaxy rotation curve.  The horizontal axis shows the
observed value of $V_{\rm flat}$ ($\approx \vmax$) for each galaxy and the vertical axis plots the value of the circular
velocity at $2$ kpc from the galaxy center.  Note that at fixed $V_{\rm flat}$,
galaxies demonstrate a huge diversity in central
densities.  Remarkably, this diversity is apparently correlated with the baryonic content in such a way as to drive the tight relation seen on the left.
The gray band in the right panel shows the expected relationship
between $\vmax$ and $V_c(2 {\rm kpc})$ for halos in \lcdm\ dark-matter-only simulations.
Clearly, the real galaxies demonstrate much more diversity than is naively predicted.

The real challenge, as we see it, is to understand how galaxies can  have
so much diversity in their rotation curve shapes compared to naive \lcdm\ expectations while
also  yielding tight correlations with baryonic content.  The fact that there is a tight correlation with {\em baryonic} mass and not stellar mass (which presumably correlates more closely with total feedback energy)  makes  the question all the more interesting.

\section{SOLUTIONS}
\label{sec:solutions}
\subsection{Solutions within \lcdm}
In this subsection, we explore some of the most popular and promising solutions
to the problems discussed above. We take as our starting point the basic \lcdm\
model plus reionization, i.e., we take it as a fundamental prediction of \lcdm\
that the heating of the intergalactic medium to $\sim 10^4\,{\rm K}$ by cosmic
reionization will suppress galaxy formation in halos with virial temperatures
below $\sim10^4\,{\rm K}$ (or equivalently, with $\vvir \la 20\,\kms$) at
$z \lesssim 6$.
\subsubsection{Feedback-induced cores}
\label{subsubsec:feedback}
Many of the most advanced hydrodynamic simulations today have shown that it is
possible for baryonic feedback to erase the central cusps shown in the density profiles in Figure 3 and produce core-like density profiles as inferred from rotation curves such as those shown in Figure 9 \citep{mashchenko2008, Pontzen2012, madau2014a, onorbe2015,read2016a}.
One key prediction is that the effect of core creation will vary with the mass
in stars formed \citep{governato2012,di-cintio2014}. If galaxies form enough
stars, there will be enough supernovae energy to redistribute dark matter and
create significant cores. If too many baryons end up in stars, however, the
excess central mass can compensate and drag dark matter back in. At the other
extreme, if too few stars are formed, there will not be enough energy in supernovae
to alter halo density structure and the resultant dark matter distribution will
resemble dark-matter-only simulations. While the possible importance of
supernova-driven blowouts for the central dark matter structure of dwarf
galaxies was already appreciated by
\citet{navarro1996a} and \citet{gnedin2002}, an important recent development is
the understanding that even low-level star formation over an extended period can
drive gravitational potential fluctuations that lead to dark matter core
formation.

\begin{figure}[t!]
\includegraphics[width=\textwidth]{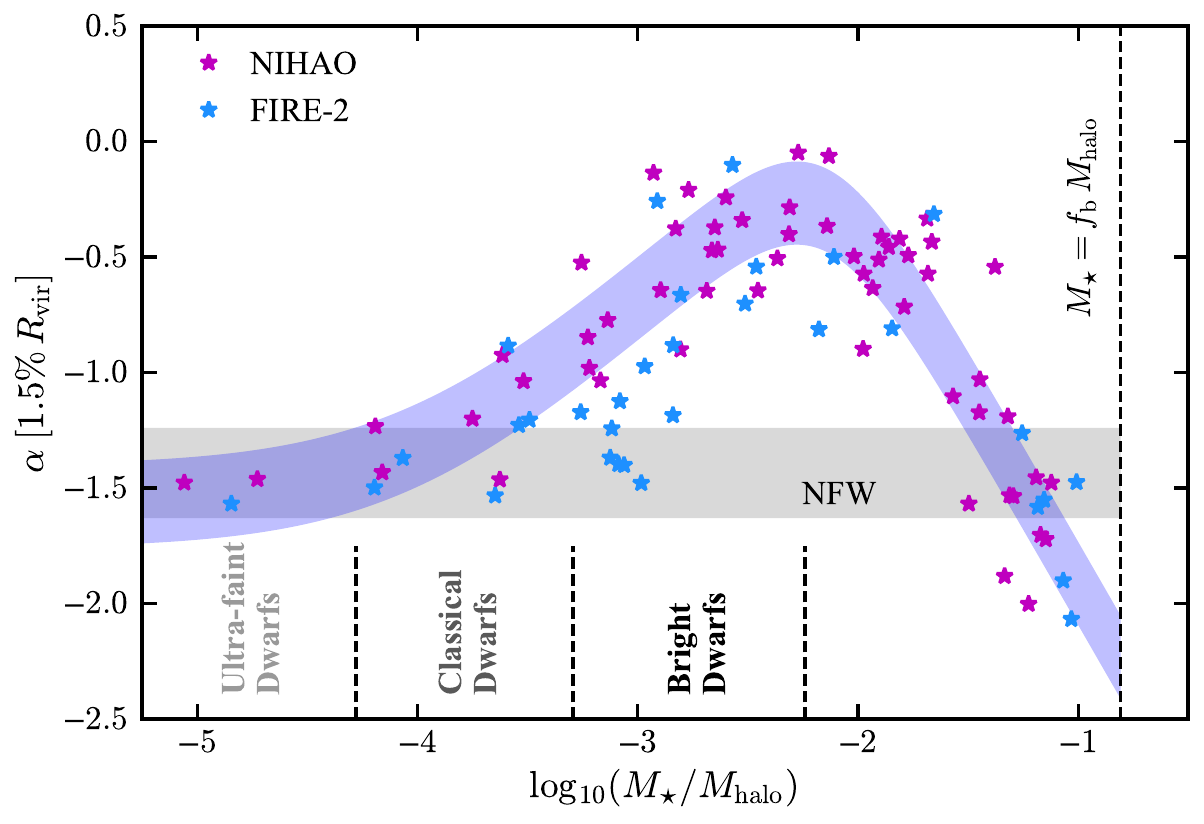}
\caption{The impact of baryonic feedback on the inner profiles of dark matter
  halos.  Plotted is the inner dark matter density slope $\alpha$ 
  at $r = 0.015 \rvir$ as a function of $M_\star/\mvir$ for simulated galaxies
  at z = 0. Larger values of $\alpha \approx 0$ imply core profiles, while lower values of $\alpha \lesssim 0.8$ imply cusps.  The shaded gray band shows the expected range of dark matter profile slopes for NFW profiles as derived from dark-matter-only simulations (including concentration scatter).  The filled magenta stars and shaded purple band (to guide the eye) show the predicted inner density slopes from the NIHAO cosmological hydrodynamic simulations by \citet{tollet2016}. The cyan stars are a similar prediction from an entirely different suite of simulations from the FIRE-2 simulations \citep[][Chan et al., in preparation]{fitts2016, hopkins2017}. Note that at dark matter core formation peaks in efficiency at $\mstar/\mvir \approx 0.005$, in the regime of the brightest dwarfs.  Both simulations find that for $\mstar/\mvir \lesssim 10^{-4}$, the impact of baryonic feedback is negligible. This critical ratio below which core formation via stellar feedback is difficult corresponds to the regime of classical dwarfs and ultra-faint dwarfs.}
\label{fig:feedback}
\end{figure}

This general behavior is illustrated in Figure \ref{fig:feedback}, which shows
the impact of baryonic feedback on the inner slopes of dark matter halos $\alpha$
measured at $1-2\%$ of the halo virial radii. Core-like density profiles have $\alpha \rightarrow 0$.  The magenta stars show results from
the NIHAO hydrodynamic simulations as a function of $\mstar/\mvir$, the ratio of
stellar mass to the total halo mass \citep{tollet2016}.  The cyan stars show 
results from an entirely different set of simulations from the FIRE-2 collaboration \citep[][Chan et al., in preparation]{wetzel2016,fitts2016,garrison-kimmel2017a}. 
 The shaded gray band shows the expected slopes 
for NFW halos with the predicted range of concentrations from dark-matter-only simulations.
We see that both sets of simulations find core formation to be fairly efficient 
$\mstar/\mvir \approx 0.005$.  This ``peak
core formation" ratio maps to $\mstar \simeq 10^{8-9} \,\msun$, corresponding to the
brightest dwarfs.  At ratios below $\mstar/\mvir \approx 10^{-4}$, however, the
impact of baryonic feedback is negligible. The ratio
below which core formation is difficult corresponds to
$\mstar \approx 10^{6} \msun$ --  the
mass-range of interest for the too-big-to-fail problem.

\begin{figure}[tb!]
\includegraphics[width=\textwidth]{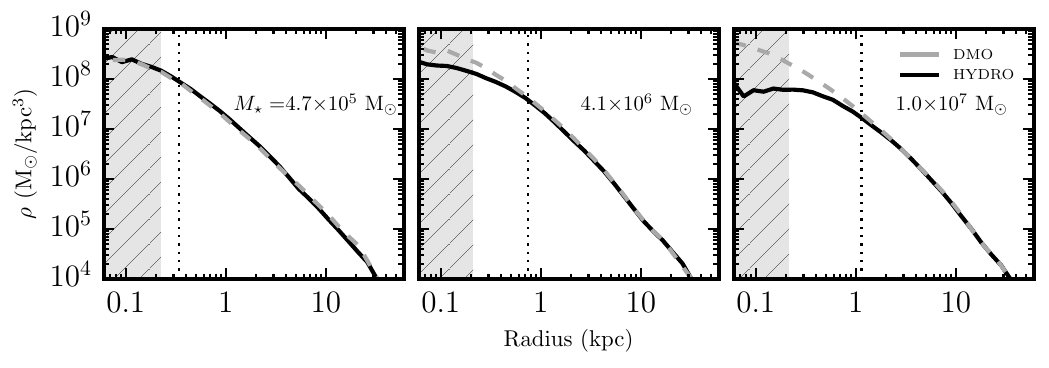}
\caption{Dark matter density profiles from full hydrodynamic FIRE-2
  simulations \citep{fitts2016}.  Shown are three different
  galaxy halos, each at mass $\mvir \approx 10^{10} \msun$.  Solid lines show
  the hydro runs while the dashed show the same halos run with dark matter only.
  The hatched band at the left of each panel marks the region where numerical
  relaxation may artificially modify density profiles and the vertical dotted line shows
  the half-light radius of the galaxy formed.  The stellar mass of the galaxy
  formed increases from left to right: $\mstar \approx 5\times10^5$,
  $4 \times 10^6$, and $10^7 \msun$, respectively.  As $\mstar$ increases, so
  does the effect of feedback. The smallest galaxy has no effect on the density
  structure of its host halo. }
\label{fig:alex}
\end{figure}

The effect of feedback on density profile shapes as a function of stellar mass
is further illustrated in Figure~\ref{fig:alex}.  Here we show simulation
results from \citet{fitts2016} for three galaxies (from a
cosmological sample of fourteen), all formed in halos with
$\mvir(z=0) \approx 10^{10} \msun$ using the FIRE-2 galaxy formation prescriptions
(\citealt{hopkins2014} and in preparation).  The dark matter density profiles of the resultant
hydrodynamical runs are shown as solid black lines in each panel, with stellar
mass labeled, increasing from left to right.  The dashed lines in each panel
show dark-matter-only versions of the same halos.  We see that only in the run that
forms the most stars ($\mstar \approx 10^7 \msun$,
$\mstar/\mvir \approx 10^{-3}$) does the feedback produce a large core.
Being conservative, for systems with $\mstar/\mvir \lesssim 10^{-4}$, feedback is likely to be ineffective in altering dark matter profiles significantly compared to dark-matter-only simulations.

\begin{summary}[SCALE WHERE FEEDBACK BECOMES INEFFECTIVE IN PRODUCING CORES]
\begin{equation*}
  \label{eq:10}
  \mstar/\mvir \approx 10^{-4} \leftrightarrow \mstar \approx 10^6 \msun 
  \leftrightarrow \mvir \approx 10^{10} \msun
  \end{equation*}
\end{summary}

It is important to note that while many independent groups are now obtaining
similar results in cosmological simulations of dwarf galaxies
\citep{governato2012, munshi2013, madau2014a, chan2015, onorbe2015, tollet2016,fitts2016} --
indicating a threshold mass of $\mstar \sim 10^6\,\msun$ or
$\mvir \sim 10^{10}\,\msun$ -- this is \textit{not} an ab initio \lcdm\
prediction, and it depends on various adopted parameters in galaxy formation
modeling. For example, \citet{sawala2016a} do not obtain cores in their
simulations of dwarf galaxies, yet they still produce systems that match observations
well owing to a combination of feedback effects that lower central densities of
satellites (thereby avoiding the \tbtf\ problem). On the other hand,
the very high resolution, non-cosmological simulations presented in
\cite{read2016a} produce cores in galaxies of \textit{all} stellar masses. We
note that Read et al.'s galaxies have somewhat higher $\mstar$ at a given
$\mvir$ than the cosmological runs described cited above; this leads to
additional feedback energy per unit dark matter mass, likely explaining the
differences with cosmological simulations and pointing to a testable prediction
for dwarf galaxies' $\mstar/\mvir$.

\subsubsection{Resolving too-big-to-fail}
\label{subsubsec:expl_tbtf}
 The baryon-induced cores described in Section~\ref{subsubsec:feedback} have their
origins in stellar feedback. The existence of such cores for galaxies above the
critical mass scale of $\mstar \approx 10^{6}\,\msun$ would explain why
$\sim$half of the classical Milky Way dwarfs -- those above this mass -- have
low observed densities. However, about half of the MW's classical dwarfs have
$\mstar < 10^{6}\,\msun$, so the scenario described in
Section~\ref{subsubsec:feedback} does not explain these systems' low central
masses. Several other mechanisms exist to reconcile \lcdm\ with the internal
structure of low-mass halos, however.

Interactions between satellites and the Milky Way -- tidal stripping, disk
shocking, and ram pressure stripping -- all act as additional forms of feedback
that can reduce the central masses of satellites. Many numerical simulations of
galaxy formation point to the importance of such interactions (which are
generally absent in dark-matter-only simulations\footnote{We note that capturing
  these effects is extremely demanding numerically, and it is not clear that any
  published cosmological hydrodynamical simulation of a Milky Way-size system 
  can resolve the mass within $300-500$ pc of satellite galaxies with the
  accuracy required to address this issue.}), and these environmental influences
are often invoked in explaining \tbtf\ (e.g., \citealt{zolotov2012,
  arraki2014, brooks2014, brook2015, wetzel2016, tomozeiu2016, sawala2016a, dutton2016}). In
many of these papers, environmental effects are limited to 1-2 virial radii from
the host galaxy. Several Local Group galaxies reside at greater distances.  While
only a handful of these systems have $\mstar < 10^6 \msun$ (most are $\mstar \sim 10^7 \msun$), these galaxies provide an initial test of the importance of external feedback: if
environmental factors are key in setting the central densities of low-mass
systems, satellites should differ systematically from field galaxies. The
results of \citet{kirby2014} find no such difference; further progress will
likely have to await the discovery of fainter systems an larger optical telescopes to provide spectroscopic samples for performing dynamical analyses. Other forms of feedback may persist
to larger distances.  For example, \citet{benitez-llambay2013} note that ``cosmic web
stripping'' (ram pressure from large-scale filaments or pancakes) may be
important in dwarf galaxy evolution.

None of these solutions would explain \tbtf\ in isolated field
dwarfs. However, a number of factors could influence the conversion between
observed HI line widths and the underlying gravitational potential, complicating
the interpretation of systematically low densities (for a discussion of some of
these issues, see \citealt{papastergis2017}). Some examples are: (1) gas
may not have the radial extent necessary to reach the maximum of the dark matter
halo rotation curve; (2) the contribution of non-rotational support (pressure
from turbulent motions) may be non-negligible and not correctly handled; and (3)
determinations of inclinations angles of galaxies may be systematically
wrong. \citet{maccio2016} find good agreement between their simulations and the
observed abundance of field dwarf galaxies in large part because the gas
distributions in the simulated dwarfs do not extend to the peak of the dark
matter rotation curve (see also \citealt{kormendy2016} for a similar conclusion reached via different considerations). A more complete understanding of observational samples
and very careful comparisons between observations and simulations are crucial
for quantifying the magnitude of any discrepancies between observations and theory.

\subsubsection{Explaining planes}
\label{subsubsec:expl_planes}
Even prior to the \citet{ibata2013} result on the potential
rotationally-supported plane in M31, multiple groups continued to study the
observed distribution of satellite galaxies, their orbits, and the consistency
of these with \lcdm. \citet{libeskind2009} and \citet{lovell2011} argued that
the MW satellite configuration and orbital distribution are consistent with
predictions from \lcdm\ simulations, while \citet{metz2009} and
\citet{pawlowski2012} argued that evidence of a serious discrepancy had only
become stronger. A major point of disagreement was whether or not filamentary
accretion within \lcdm\ is sufficient to explain satellite orbits. Given that
SDSS only surveyed about 1/3 of the northern sky (centered on the North Galactic
Pole, thereby focusing on the portion of the sky where the polar plane was
claimed to lie), areal coverage was a serious concern when trying to
understand the significance of the polar distribution of satellites. DES has
mitigated this concern somewhat, but it is also surveying near the polar
plane. \citet{pawlowski2016} has recently argued that incomplete sky coverage is
\textit{not} the driving factor in assessing phase-space alignments in the Milky Way;
future surveys with coverage nearer the Galactic plane should definitively test
this assertion.

Following \citet{ibata2013}, the question of whether the M31 configuration (right-hand side of Figure 11) is
expected in \lcdm\ also became a topic of substantial interest. The general
consensus of work rooted in \lcdm\ is that planes qualitatively similar (though not as thin) as the M31 plane are
not particularly uncommon in \lcdm\ simulations, but that these planes are not
rotationally-supported structures (e.g., \citealt{bahl2014, gillet2015,
  buck2016}). Since we
view the M31 plane almost perfectly edge-on, proper motions of dwarf galaxies in
the plane would provide a clean test of its nature. Should this plane turn out
to be rotationally supported, it would be \textit{extremely} difficult to
explain with our current understanding of the \lcdm\ model. These proper motions
may be possible with a combination of \textit{Hubble} and \textit{James Webb
  Space Telescope} data. \citet{skillman2017} presented preliminary observations
of three plane and three non-plane galaxies, finding no obvious differences
between the two sets of galaxies. Future observations of this sort could help
shed light on the M31 plane and its nature.

\subsubsection{Explaining the radial acceleration relation}
\label{subsubsec:expl_relations}
Almost immediately after  \citet{McGaugh2016} published their RAR relation paper, \citet{keller2017} responded by demonstrating that a similar relation can be obtained
using \lcdm\ hydrodynamic simulations of disk galaxies.  Importantly, however, the systems
simulated did not include low-mass galaxies, which are dark-matter-dominated throughout.  The smallest galaxies are the ones with low acceleration in their centers as well as in their outer parts, and they remain the most puzzling to explain (see \citealt{Milgrom2016} for a discussion related to this issue).  

More recently, \citet{Navarro2016} have argued that  \lcdm\ can naturally produce an acceleration relation similar to that shown in Figure 12.   A particularly compelling section of their argument follows directly from abundance-matching (Figure 6): the most massive disk galaxies that exist are not expected to be in halos much larger than  $5 \times 10^{12} \msun$.  This sets a maximum acceleration scale ($\sim 10^{-10}$ m s$^{-2}$) above which any observed acceleration {\it must} track the baryonic acceleration.  The implication is that any mass-discrepancy attributable to dark matter will only begin to appear at accelerations below this scale.  Stated slightly differently, any successful model of galaxy formation set within a \lcdm\ context \textit{must} produce a relation that begins to peel above the one-to-one only below the characteristic scale observed.  

It remains to be seen whether the absolute normalization and shape of the RAR in the low-acceleration regime can be reproduced in \lcdm\ simulations that span the full range of galaxy types that are observed to obey the RAR.  As stated previously, these same simulations must also simultaneously reproduce the observed diversity of galaxies at fixed $\vmax$ that is seen in the data (e.g., as shown in the right-panel of Figure 12).

\subsection{Solutions requiring modifications to \lcdm }
\subsubsection{Modifying linear theory predictions}
\label{subsubsec:linear}
As discussed in Section~\ref{subsec:particle_physics}, the dominant impact of dark matter particle nature on the linear theory power spectrum for CDM models is in the high-$k$ cut-off (see labeled curves in Figure~\ref{fig:pofk}). This cut-off is set by the
free-streaming or collisional damping scale associated with CDM and is of order 1 comoving pc
(corresponding to perturbations of $10^{-6}\,\msun$) for canonical WIMPs \citep{green2004} or 0.001 comoving pc (corresponding to $10^{-15}\,\msun$) for a $m\approx 10 \mu{\rm eV}$ QCD axion \citep{nambu1990}. 
In these models, the dark matter halo
hierarchy should therefore extend 18 to 27 orders of magnitude below the mass
scale of the Milky Way ($10^{12}\,\msun$; see Fig.~\ref{fig:pofk}). 

A variety of dark matter models result in a truncation of linear perturbations
at much larger masses, however. For example, WDM models have an effective free-streaming length $\lambda_{\rm fs}$ that scales inversely with particle mass \citep{bode2001, viel2005}; in the Planck \citeyearpar{planck2015} cosmology, this relation is approximately
\begin{equation}
\label{eq:lambda_fs}
\lambda_{\rm fs}=70 \, \left(\frac{m_{\rm WDM}}{1\,{\rm keV}}\right)^{-1.11}\,\kpc
\end{equation}
and the corresponding free-streaming mass is 
\begin{eqnarray}
\label{eq:m_fs}
M_{\rm fs}&=&\frac{4\pi}{3}\,\rho_{\rm m}\,\left(\frac{\lambda_{\rm fs}}{2}\right)^{3}\nonumber\\
&=&7.1\times 10^6\,\left(\frac{m_{\rm WDM}}{1\,{\rm keV}} \right)^{-3.33}\,\msun .
\end{eqnarray}
The effects of
power spectrum truncation are not limited to the free-streaming scale, however:
power is substantially suppressed for significantly larger scales (smaller
wavenumbers $k$). A characterization of the scale at which power is
significantly affected is given by the half-mode scale $k_{\rm
  hm}=2\,\pi/\lambda_{\rm hm}$, where the transfer function is reduced by 50\%
relative to CDM. The half-mode wavelength $\lambda_{\rm hm}$ is approximately fourteen times larger than the free-streaming length \citep{schneider2012}, meaning that structure below $\sim 5 \times 10^{10}\,\msun$ is significantly different from CDM in a 1 keV thermal dark matter model:
\begin{equation}
M_{\rm hm}=1.9\times 10^{10} \, \left(\frac{m_{\rm WDM}}{1\,{\rm keV}} \right)^{-3.33}\,\msun\,.
\end{equation}
Examples of power suppression  for several thermal WDM models are shown by the dashed, dotted, and dash-dotted lines in Fig. \ref{fig:pofk}.

The lack of small-scale power in models with warm (or hot) dark matter is a
testable prediction. As the free-streaming length is increased and higher-mass
dark matter substructure is erased, the expected number of dark matter
satellites inside of a Milky Way-mass halo decreases. The observed number of
dark-matter-dominated satellites sets a lower limit on the number of subhalos
within the Milky Way, and therefore, a lower limit on the warm dark matter
particle mass. \citet{polisensky2011} find this constraint is $m>2.3\,{\rm keV}$
(95\% confidence) while \citet{lovell2014} find $m>1.6\,{\rm keV}$; these
differences come from slightly different cosmologies, assumptions about the
mass of the Milky Way's dark matter halo, and modeling of completeness limits
for satellite detections.

It is important to note that particle mass and the free-streaming scale are
\textit{not} uniquely related: the free-streaming scale depends on the particle
production mechanism and is set by the momentum distribution of the dark matter
particles. For example, a resonantly-produced sterile neutrino can have a much
``cooler'' momentum distribution than a particle of the same mass that is
produced by a process in thermal equilibrium \citep{shi1999}. Constraints therefore must be
computed separately for each production mechanism \citep{merle2015, venumadhav2016}. As an example, the effects of Dodelson-Widrow \citeyearpar{dodelson1994} sterile neutrinos, which are produced through non-resonant oscillations from active neutrinos, can be matched to effects of thermal relics via the following relation:
\begin{equation}
\label{eq:viel-bozek}
m(\nu_s)=3.9\,{\rm keV}\,\left(\frac{m_{\rm thermal}}{1\,{\rm keV}}\right)^{1.294}\,\left(\frac{\Omega_{\rm DM}h^2}{0.1225}\right)^{-1/3}
\end{equation}
\citep{abazajian2006,bozek2016}.

\begin{figure}
 \begin{minipage}{0.325\textwidth}
 \includegraphics[width=0.99\linewidth]{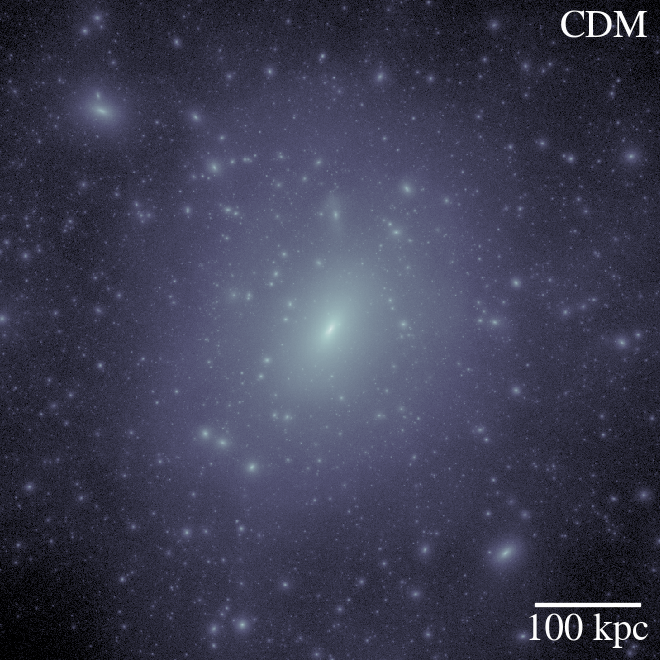}
 \end{minipage}
 \begin{minipage}{0.325\textwidth}
 \includegraphics[width=0.99\linewidth]{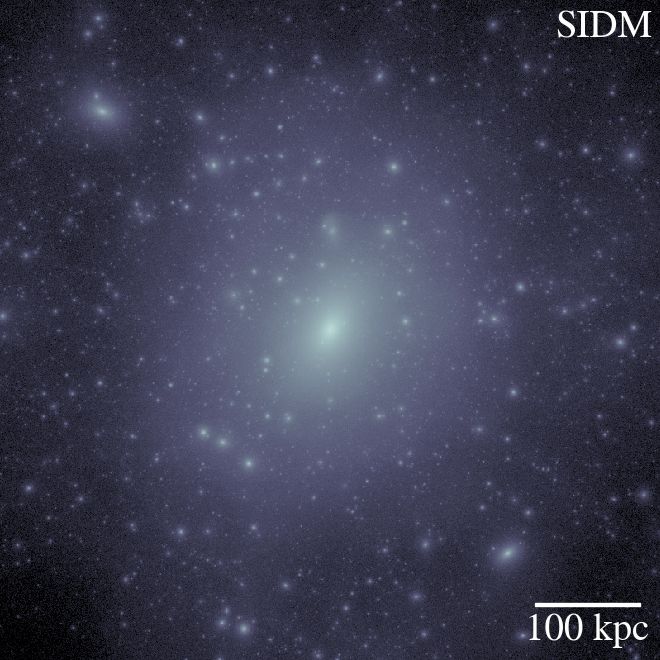}
 \end{minipage}
 \begin{minipage}{0.325\textwidth}
 \includegraphics[width=0.99\linewidth]{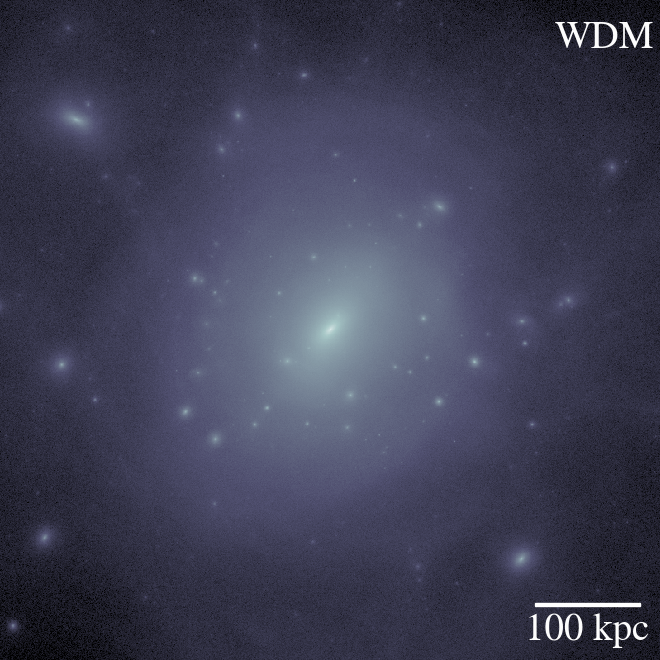}
 \end{minipage}
 \begin{minipage}{0.325\textwidth}
 \includegraphics[width=0.99\linewidth]{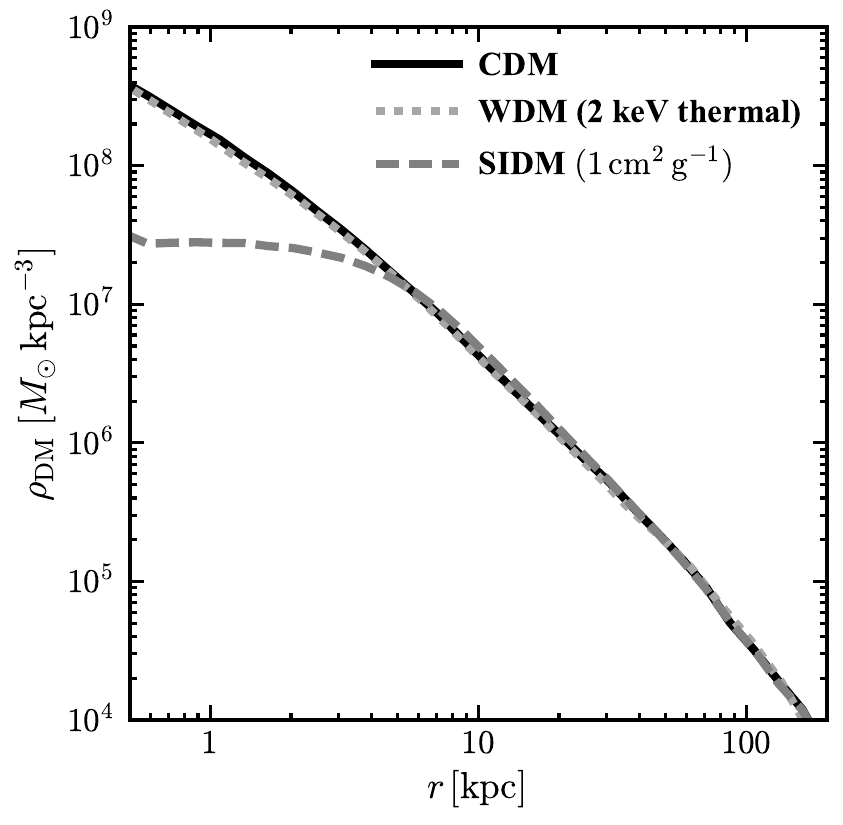}
 \end{minipage}
 \begin{minipage}{0.325\textwidth}
\includegraphics[width=0.99\linewidth]{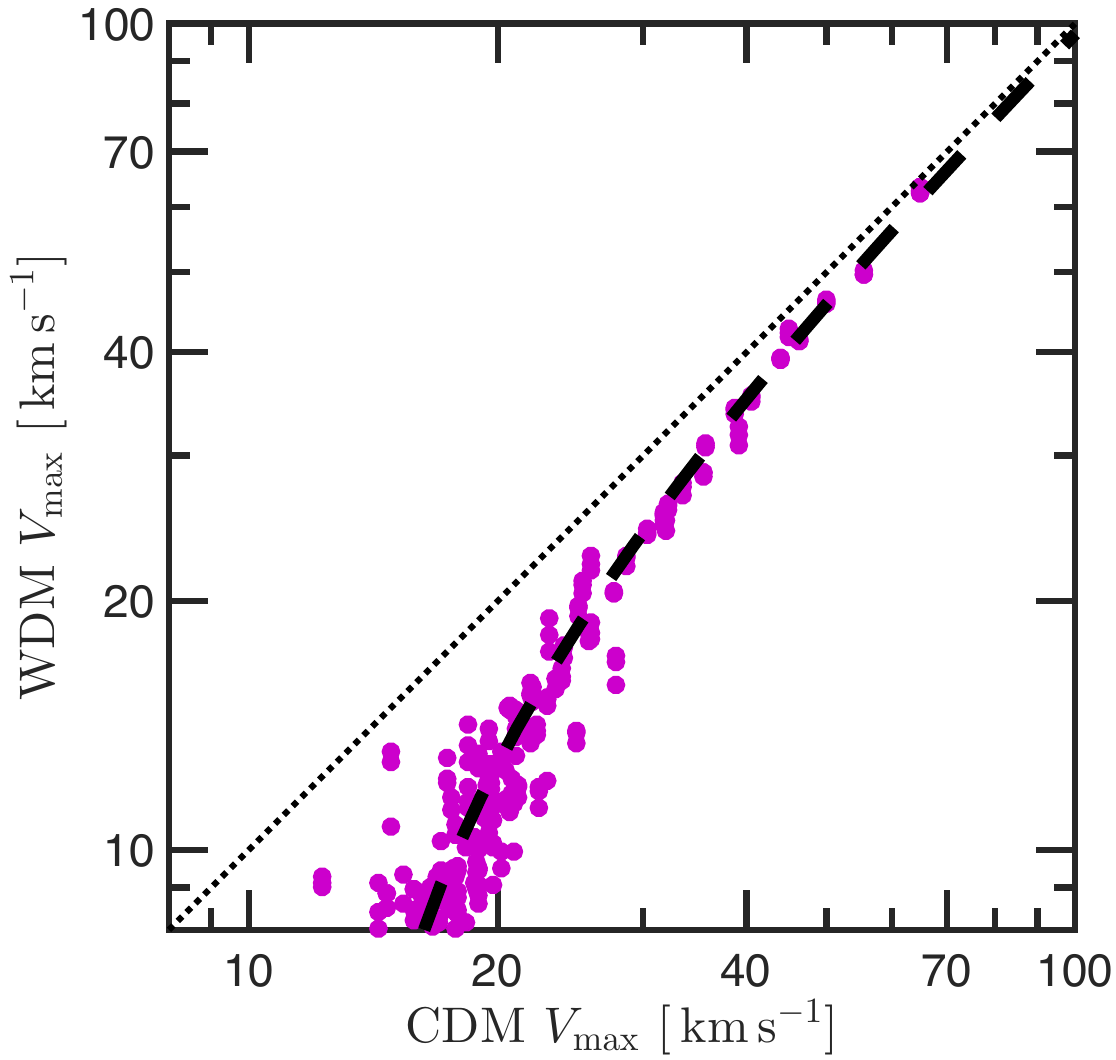}
\end{minipage}
\begin{minipage}{0.325\textwidth}
\includegraphics[width=\linewidth]{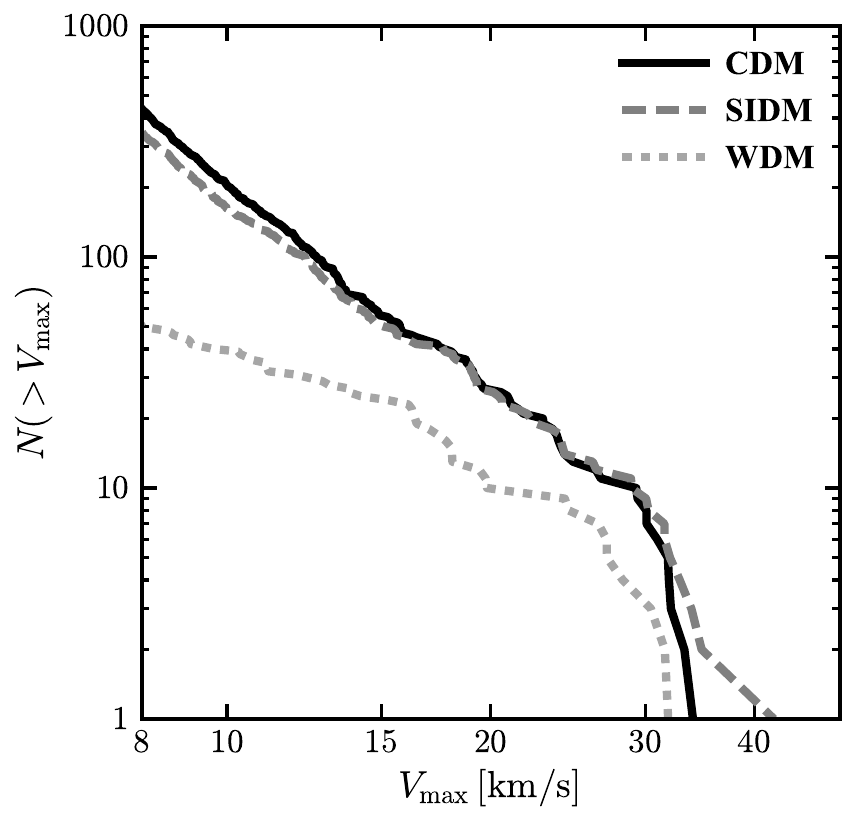}
\end{minipage}
\caption{Dark matter phenomenology in the halo of the Milky Way.  The three images in the upper row show the same Milky-Way-size dark matter halo simulated with
CDM, SIDM ($\sigma/m = 1~{\rm cm^2/g}$), and WDM (a Shi-Fuller resonant model with a thermal equivalent mass of $2~{\rm keV}$).  The left panel in the bottom row shows the dark matter density profiles of the same three halos while the bottom-right panel shows the subhalo velocity functions for each as well. The middle panel on the bottom shows that while the host halos have virtually identical density structure in WDM and CDM, individual subhalos identified in both simulations smaller $\vmax$ values in WDM \citep{bozek2016}.  This effect can explain the bulk of the differences seen in the $\vmax$ functions (bottom right panel).  Note that SIDM does not reduce the abundance of substructure (unless the power spectrum is truncated) but it does naturally produce large constant-density cores in the dark matter distribution.  WDM does not produce large constant-density cores at Milky Way-mass scales but does result in fewer subhalos near the free-streaming mass and reduces $\vmax$ of a given subhalo (through reduced concentration) near the half-mode mass ($\mhalo \la 10^{10}\,\msun$ for the plotted 2 keV thermal equivalent model).
}
\label{fig:non-cdm-images}
\end{figure}

The effects of power spectrum suppression are not limited to pure number counts of dark matter halos: since cosmological structure form hierarchically, the erasure of small perturbations affects the collapse of more massive objects. The primary result of this effect is to delay the assembly of halos of a given mass relative to the case of no power spectrum suppression. Since the central densities of dark matter halos reflect the density of the Universe at the time of their formation, models with reduced small-scale power also result in shallower central gravitational potentials at fixed total mass for halos within 2-3 dex of the free-streaming mass. This effect is highlighted in the lower-middle panel of Figure~\ref{fig:non-cdm-images}. It compares $\vmax$ values for a CDM simulation and a WDM simulation that assumes a thermal-equivalent mass of 2 keV but is otherwise identical to the CDM simulation. Massive halos ($\vmax \ga 50 \,\kms$) have identical structure; at lower masses, WDM halos have systematically lower $\vmax$ values than their CDM counterparts. This effect comes from a reduction of $\vmax$ for a given halo in the WDM runs, {\em not} from there being fewer objects. The reduction in central density due to power spectrum suppression for halos near or just below the half-mode mass (but significantly more massive than the free-streaming mass) is how WDM can solve the \tbtf\ problem \citep{anderhalden2013}.

\subsubsection{Modifying non-linear predictions}
\label{subsubsec:nonlinear}
The non-linear evolution of CDM is described by the Collisionless Boltzmann
equation. Gravitational interactions are the only ones that are relevant for CDM
particles, and these interactions operate in the mean field limit (that is,
gravitational interactions between individual DM particles are negligible
compared to interactions between a dark matter particle and the large-scale
gravitational potential). The question of how strong the constraints are on
non-gravitational interactions between individual dark matter particles is
therefore crucial for evaluating non-CDM models.

There has been long-standing interest in models that involve dark matter
self-interactions \citep{carlson1992, spergel2000}. In its simplest form,
self-interacting dark matter (SIDM, sometimes called collisional dark matter) is characterized by an energy-exchange interaction cross section $\sigma$. The mean free path $\lambda$ of dark matter particles is then 
$\lambda=(n\,\sigma)^{-1}$, where $n$ is the local number density of dark matter
particles. Since the mass of the dark matter particle is not known, it is often
useful to express the mean free path as $(\rho\,\sigma/m)^{-1}$ and to quantify
self-interactions in terms of the cross section per unit particle mass,
$\sigma/m$. If $\lambda(r)/r \ll 1$ at radius $r$ from the center of a dark matter halo, many scattering events occur per local
dynamical time and SIDM acts like a fluid, with conductive transport of heat. In
the opposite regime, $\lambda(r)/r \gg 1$, particles are unlikely to scatter
over a local dynamical time and SIDM is effectively an optically thin (rarefied)
gas, with elastic scattering between dark matter particles.  Most work in recent years has
been far from the fluid limit.

As originally envisioned by \citet{spergel2000} in the context of solving the missing satellites and cusp/core problems, the mean free path for
self-interactions is of order $1\,\kpc \la \lambda \la 1\,\mpc$ at
densities characteristic of the Milky Way's dark matter halo (0.4
Gev/${\rm cm^3}$; \citealt{read2014}), leading to self-interaction cross sections of
$400 \ga \sigma/m \ga 0.4\,{\rm cm^2/g}$
($800 \ga \sigma/m \ga 0.8 \,{\rm barn/GeV}$). This scale ($\sim$barn/GeV) is
enormous in particle physics terms -- it is comparable to the cross-section for
neutron-neutron scattering -- yet it remains difficult to exclude observationally.  It is important to emphasize that the dark matter particle self-interaction strength can, in principle, be completely decoupled from the dark matter's interaction strength with Standard Model particles and thus standard direct detection constraints offer no absolute model-independent limits on $\sigma/m$ for the dark matter.  Astrophysical constraints are therefore essential for understanding dark matter physics.

Though the SIDM cross section estimates put forth
by \citet{spergel2000} were based on analytic arguments, the interaction scale they
proposed to alleviate the cusp/core problem does overlap (at the low end) with more modern results based on fully self-consistent cosmological simulations.   Several groups have now run cosmological simulations
with dark matter self-interactions and have found that  models with
$\sigma/m \approx 0.5-10 \,{\rm cm^2/g}$ produce dark matter cores in dwarf galaxies with sizes $\sim 0.3-1.5\,
\kpc$ and central densities  $2-0.2 \times 10^{8}\,\msun\,\kpc^{-3}=7.4-0.74
\,{\rm GeV\,cm^{-3}}$ that can alleviate the cusp/core and \tbtf\ problems discussed above (e.g., \citealt{vogelsberger2012,peter2013,fry2015,elbert2015}).  SIDM does not, however, significantly alleviate the missing satellites problem, as the  substructure counts in SIDM simulations are almost identical to those in CDM simulations (\citealt{Rocha2013}; see Figure~\ref{fig:non-cdm-images}).

One important constraint on possible SIDM models comes from galaxy clusters.  The high central dark matter densities observed in clusters exclude SIDM models with $\sigma/m \ga 0.5\,{\rm cm^2/g}$, though SIDM with $\sigma/m \simeq 0.1 {\rm cm^2/g}$ may be  preferred over CDM (e.g., \citealt{kaplinghat2016,Elbert2016}).  This means that in order for SIDM is to alleviate the small-scale problems that arise in standard CDM and also match constraints seen on the galaxy cluster scale, it needs to have a velocity-dependent cross section $\sigma(v)$ that decreases as the rms speed of dark matter particles involved in the scattering rises from the scale of dwarfs ($v \sim 10 ~\kms$) to  galaxy clusters ($v \sim 1000 ~\kms$).  Velocity-dependent scattering cross sections are not uncommon among Standard Model particles.
  
Figure 15 shows the results of three high-resolution cosmological simulations (performed by V. Robles, T. Kelley, and B. Bozek in collaboration with the authors) of the same Milky Way mass halo  done with  CDM, SIDM ($\sigma/m = 1 \,{\rm cm^2/g}$), and WDM (a 7 keV resonant model, with thermal-equivalent mass of 2 keV). The images show density maps spanning 600 kpc.  It is clear that while WDM produces many fewer subhalos than CDM, the SIDM model yields a subhalo distribution that is very similar to CDM, with only slightly less substructure near the halo core, which itself is slightly lower density than the CDM case. 

These visual impressions are quantified in the bottom three panels, which show the main halo density profiles (left) and the subhalo $\vmax$ functions for all three simulations (right).  The middle panel shows the relationship between the $\vmax$ values of individual halos identified in both CDM and WDM simulations \citep{bozek2016}. The left panel shows clearly that SIDM produces a large, constant-density core in the main halo, while the WDM profile is almost identical to the CDM case. However, for mass scales close to the half-mode suppression mass of the WDM model ($\mhalo \la 10^{10}\,\msun$ for this case), the density structure is affected significantly. This effect accounts for most of the difference seen in the right panel: WDM subhalos have $\vmax$ values that are greatly reduced compared to their CDM counterparts, meaning there is a $\vmax$-dependent shift \textit{leftward} at fixed number (i.e., subhalos at this mass scale are not being destroyed, which would result in a reduction in number at fixed $\vmax$).

Finally, we conclude by noting that it is possible to write down SIDM models that have both truncated power spectra and significant self-interactions.  Such models produce results that are a hybrid between traditional WDM and SIDM with scale-invariant spectra \citep[e.g.][]{cyr-racine2016, vogelsberger2016}.  Specifically, it is possible to modify dark matter in such a way that it produces both fewer subhalos (owing to power spectra effects) and constant density cores (owing to particle self-interactions) and thus solve the substructure problem and core/cusp problem simultaneously without appealing to strong baryonic feedback.

\section{Current Frontiers}
\label{sec:frontiers}
\subsection{Dwarf galaxy discovery space in the Local Group}
\label{subsec:discoveries}
The tremendous progress in identifying and characterizing faint stellar systems
in the Local Group has led to a variety of new questions. For one, these 
discoveries have blurred what was previously a clear difference between dwarf
galaxies and star clusters, leading to the question, ``what is a galaxy?''
\citep{willman2012}. DES has identified several new satellite galaxies, many of
which appear to be clustered around the Large Magellanic Cloud (LMC; \citealt{drlica-wagner2015}). The putative association of these satellites with the LMC is intriguing
\citep{jethwa2016,sales2017}, as the nearly self-similar nature of dark matter
substructure implies that the LMC -- which is likely to be hosted by a halo of
$M_{\rm peak} \sim 10^{11}\,\msun$ \citep{boylan-kolchin2010} -- could itself contain
multiple dark matter satellites above the mass threshold required for galaxy
formation. Satellites of the LMC and even fainter dwarfs will be attractive
targets for ongoing and future observations to test basic predictions of \lcdm\
\citep{wheeler2015}. 

The 800 pound gorilla in the dwarf discovery landscape is the Large Synoptic
Survey Telescope (LSST). Currently under construction and set to begin
operations in 2022, LSST has the potential to expand dwarf galaxy
discovery space substantially: by the end of the survey, co-added LSST data will
be sensitive to galaxies ten times more distant (at fixed luminosity) than SDSS,
or equivalently, LSST will be able to detect galaxies that are one hundred times
fainter than SDSS at the same distance. This means that LSST should be complete
for galaxies with $L_\star \ga 2 \times 10^3\,\lsun$ within $\sim 1\,\mpc$ of the
Galaxy, dramatically increasing the census of very faint galaxies beyond $\sim
100\,\kpc$ from the Earth. 

One of the unique features of LSST data sets will be the ability to explore the
properties of low-mass, \textit{isolated} dark matter halos (i.e., those that
have not interacted with a more massive system such as the Milky Way), thereby
separating out the effects of environment from internal feedback and dark matter
physics. Given the predictions discussed in Sec.~\ref{subsubsec:feedback}, any
new discoveries with $\mstar \la 10^{6}\,\msun$ at $\sim 1\,\mpc$ from the Milky
Way and M31 will be attractive targets for discriminating between baryonic
feedback and dark matter physics. At this distance, spectrographs on 10m-class
telescopes will not be sufficient to measure kinematics of resolved stars;
planned 30m-class telescopes will be uniquely suited to this task.

In addition to hosting surviving satellites, galactic halos also act as a
graveyard for satellite galaxies that have been disrupted through tidal
interactions. These disrupted satellites can form long-lived tidal streams; more
generally, the stars from these satellites are part of a galaxy's stellar halo
(which may also encompass stars from globular clusters or other
sources). Efforts are underway to disentangle disrupted satellites from other
stars in the Milky Way halo via chemistry and kinematics (see \citealt{bland-hawthorn2016} for a recent review).

\subsection{Dwarfs beyond the Local Group}
An alternate avenue to probing deeper within the Local Group is to search for
low-mass galaxies further away (but still in the very local Universe). The Dark
Energy Camera (DECam) and Subaru/Hyper Suprime-Cam are being used by several
groups to search for very faint companions in a variety of systems (from NGC
3109, itself a dwarf galaxy at $\sim 1.3\,\mpc$, to Centaurus A, a relatively
massive elliptical galaxy at $\approx 3.8\,\mpc$ (\citealt{sand2015a,
  crnojevic2016, carlin2016}). Searches for the \textit{gaseous} components of
galaxies that would otherwise be missed by surveys have also proven fruitful,
with a number of individual discoveries \citep{giovanelli2013, sand2015,
  tollerud2016}. 

Recently, the rediscovery of ultra-diffuse dwarf galaxies \citep{impey1988,
  dalcanton1997, koda2015, van-dokkum2015} has led to significant interest in
these odd systems, which have sizes comparable to the Milky Way but luminosities
comparable to bright dwarf galaxies. Ultra-diffuse dwarfs have been discovered
predominantly in galaxy clusters, but if similar systems -- perhaps with even
lower luminosities -- exist near the Local Group, they could have escaped
detection. Understanding the formation and evolution of ultra-diffuse dwarfs, as
well as their dark matter content and connection to the broader galaxy
population, has the potential to alter our current understanding of faint
stellar systems.

\subsection{Searches for starless dwarfs}
Very low mass dark matter halos \textit{must} be starless, should they
exist. Detecting starless halos would represent a strong confirmation of the
\lcdm\ model (and would place stringent constraints on the possible solutions to problems
covered in this review); accordingly, astronomers and physicists are exploring a
variety of possibilities for detecting such halos.

A promising technique for inferring the presence of the predicted population of
low-mass, dark substructure within the Milky Way is through subhalos' effects on
very cold low velocity dispersion stellar streams \citep{ibata2002,
  carlberg2009a, yoon2011}. Dark matter substructure passing through a stream will
perturb the orbits of the stars, creating gaps and bunches in the
stream. Although many physical phenomena may produce similar effects, and the
very existence of gaps themselves remains a matter of debate, large samples of
cold streams would likely provide the means to test the abundance of low-mass
($M_{\rm vir} \sim 10^{5-6}\,\msun$) substructure in the Milky Way. We note that the
streams from disrupting satellite galaxies discussed above are not suitable for
this technique, as they are produced with large enough stellar velocity
dispersions that subhalos' effects will go undetectable. Blind surveys for HI gas provide yet another path to searching for starless (or
extremely faint) substructure in the very nearby Universe. Some ultra-compact
high-velocity clouds (UCHVCs) may be gas-bearing ``mini-halos'' that are devoid
of stars (e.g., \citealt{blitz1999}).

Most of the probes we have discussed so far rely on electromagnetic signatures
of dark matter. Gravitational lensing is unique in that it is sensitive to
\textit{mass} alone, potentially providing a different window into low-mass dark
matter halos.  
\citet{vegetti2010, vegetti2012} have detected two relatively low-mass dark
matter subhalos within lensed galaxies using this technique. The galaxies are at
cosmological distances, making it difficult to identify any stellar component
associated with the subhalos; Vegetti et al. quote upper limits on the
luminosities of detected subhalos of $\sim 5\times 10^{6-7}\,\lsun$, comparable
to classical dwarfs in the Local Group. The inferred dynamical masses are much
higher, however: within 300 pc, Milky Way satellites all have $M_{300} \approx
10^7\,\msun$ \citep{strigari2008}, while the detected subhalos have $M_{300} \approx (1-10)\times
10^{8}\,\msun$. It remains to be seen whether this is related to the lens
modeling or if the substructure in lensing galaxies is fundamentally different
from that in the Local Group.

More recently, ALMA has emerged as a promising tool for detecting dark matter
halo substructure via spatially-resolved spectroscopy of lensed
galaxies. This technique was discussed in \citet{hezaveh2013}, and recently, a
subhalo with a total mass of $\sim 10^{9}\,\msun$ within $\sim 1\,\kpc$ was
detected with ALMA \citep{hezaveh2016}. At present, the detected substructure is significantly more massive than the
hosts of dwarf galaxies in the Local Group: the velocity dispersion of the
substructure is $\sigma_{\rm DM} \sim 30\,\kms$ as opposed to
$\sigma_{\star} \approx 5-10\,\kms$ for Local Group dwarf satellites. This value
of $\sigma_{\rm DM}$ is indicative of a galaxy similar to the Small Magellanic
Cloud, which has $\mstar \sim 5\times 10^8\,\msun$ and
$\mvir \sim (5-10) \times 10^{10}\,\msun$. The discovery of
additional lens systems, and the enhanced resolution and sensitivity of ALMA in
its completed configuration, promise to reveal lower-mass substructure, perhaps
down to scales similar to Local Group satellites but at cosmological distances
and in very different host galaxies.

\subsection{Indirect signatures of dark matter}
If dark matter is indeed a standard WIMP, two dark matter particles can
annihilate into Standard Model particles with electromagnetic signatures. This
process is exceedingly rare, on average; as discussed in
Section~\ref{subsec:particle_physics}, the freeze-out of dark matter
annihilations is what sets the relic density of dark matter in the WIMP
paradigm. Nevertheless, the annihilation rate is proportional to the local value
$\rho_{\rm DM}^2$, meaning that the centers of dark matter halos are potential
sites for annihilations. While the brightest source of such annihilations in the
sky should be the Galactic Center, foregrounds make unambiguous detection of
annihilating dark matter toward the Galaxy challenging. Dwarf
spheroidal galaxies have somewhat lower predicted annihilation fluxes owing both
to their greater distances and lower masses, but they have the significant
advantage of being free of foreground contamination. The {\it Fermi}
$\gamma$-ray telescope has surveyed MW dwarfs extensively, with no conclusive
evidence for dark matter annihilation products. The upper limits on combined
dwarf data from {\it Fermi} are already placing moderate tension on the most
basic ``WIMP miracle'' predictions for the annihilation cross section for wimps
with $m \la 100\,{\rm GeV}$ \citep{ackermann2015}. Searches for annihilation from starless dark matter subhalos within the Milky Way via the \textit{Fermi} point source catalog have not
yielded any detections to date \citep{calore2016}.

On cosmic scales, dark matter annihilations may contribute to the extragalactic
gamma-ray background \citep{zavala2010}. The expected contributions of dark matter depend
sensitively on the spectrum of dark matter halos and subhalos, as well as
the relation between concentration and mass for very low mass systems. These
relations can be estimated by a variety of methods (though generally not
simulated directly, owing to the enormous range of scales that contribute), with
uncertainties being grouped into a ``boost factor'' that describes unresolved
annihilations. 

If dark matter is a sterile neutrino rather than a WIMP-like particle,
self-annihilation will not be seen. Sterile neutrinos decay radiatively to an
active neutrino and a photon, however; for all of the relevant sterile neutrino
parameter space, this decay is effectively at rest and a clean signature is
therefore a spectral line at half the rest mass energy of the dark matter
particle, $E_{\gamma}=m_{\rm DM}/2$. While there is no \textit{a priori}
expectation for the mass of the sterile neutrino, arguments from
Section~\ref{subsubsec:linear} point to $E_{\gamma} \ga 1$ keV, so searches in
the soft X-ray band are constraining. The most promising recent result in this
field is the detection of a previously unknown X-ray line near 3.51 keV in the
spectra of individual galaxy clusters, stacked galaxy clusters, and the halo of
M31 \citep{bulbul2014, boyarsky2014}. X-ray observations and satellite counts in
M31 rule out an oscillation (\citealt{dodelson1994}) origin for this line if it
indeed originates from sterile neutrino dark matter \citep{horiuchi2014},
leaving heavy scalar decay and possibly resonant conversion as possible
production mechanisms \citep{merle2015}. A definitive test of the origin of the
3.5 keV line was expected from the \textit{Hitomi} satellite, as it had the
requisite energy resolution to see the thermal broadening of the line due to
virial motions (i.e., the line width from a halo with mass $\mvir$ should be
$\sim \vvir/c$). With \textit{Hitomi}'s untimely demise, tests of the line's
origin may have to wait for \textit{Athena}.

\subsection{The high-redshift Universe}
While studies of low-mass dark matter halos are most easily conducted in the
very nearby Universe owing to the faintness of the galaxies they host, there are
avenues at higher redshifts that may provide alternate windows in to the
spectrum of density perturbations. One potentially powerful probe at $z\sim 2-6$
is the \textit{Lyman-$\alpha$ forest} of absorption lines produced by neutral
hydrogen in the intergalactic medium between us and high-redshift quasars (see
\citealt{mcquinn2016} for a recent review and further details). This hydrogen
probes the density field in the quasi-linear regime (i.e., it is in
perturbations that are just starting to collapse) and can constrain the dark
matter power spectrum to wavenumbers as large as $k \sim 10\,h\,\mpc^{-1}$. Any
model that reduces the power on this scale relative to \lcdm\ expectations will
predict different absorption patterns. In particular, WDM will suppress power on
these scales.

\citet{viel2013} used Lyman-$\alpha$ flux power spectra from 25 quasar
sightlines to constrain the mass of thermal relic WDM particle to $m_{\rm WDM,
  th} > 3.3\,{\rm keV}$ at 95\% confidence. This translates into a density
perturbation spectrum that must be very close to \lcdm\ down to $M \sim
10^{8}\,\msun$ \citep{schneider2012} and would rule out the possibility that
free-streaming has direct relevance for the scales of classical dwarfs (and
larger-mass systems). The potential complication with this interpretation is the
relationship between density and temperature in the intergalactic medium, as
pressure or thermal motions can mimic the effects of dark matter
free-streaming.

Counts of galaxies in the high-redshift Universe also trace the spectrum of
collapsed density perturbations at low masses, albeit in a non-trivial
manner. The mere existence of galaxies at high redshift places an upper limit on
the free-streaming length of dark matter (so long as all galaxies form within
dark matter halos) in much the same way that the existence of substructure in
the local Universe does \citep{schultz2014}. \citet{menci2016} have placed limits on the masses of
thermal relic WDM particles of 2.4 keV (2.1 keV) at $68\%$ (95\%) confidence
based on the detection of a single galaxy in the
\textit{Hubble} Frontier Fields at $z \sim 6$ with absolute UV magnitude of
$M_{\rm UV}=-12.5$ \citep{livermore2017}. While this stated constraint is very strong, and the
technique is promising, correctly modeling faint, high-redshift galaxies --
particularly lensed ones -- at can be very challenging. Furthermore, the true
redshift of the galaxy can only be localized to $\Delta z \sim 1$; the rapid
evolution of the halo mass function at high redshift further complicates
constraints. With the upcoming \textit{James Webb Space Telescope}, the
high-redshift frontier will be pushed fainter and to higher redshifts, raising
the possibility of placing strong constraints on the free-streaming length of
dark matter through structures in the early Universe.

\begin{textbox}[ht]\section{The challenge of detecting ``empty'' dark matter
    halos }
The detection of abundant, baryon-free, low-mass dark matter halos would be an
unambiguous validation of the particle dark matter paradigm, would strongly constrain
particle physics models, and would eliminate many of the dark matter candidates
for the origin of the small-scale issues described in this review. Why is this
such a challenging task? \\

The answer lies in the densities of low-mass dark
matter halos compared to other astrophysical objects. From
Equation~\eqref{eq:mvir}, the average density within a halo's virial radius is
200 times the cosmic matter density. For the most abundant low-mass halos in
standard WIMP models -- those just above the free-streaming scale of
$\sim 10^{-6}\,\msun$ -- the virial radius is approximately $0.1\,{\rm
  pc}$.
This is the equivalent of the mass of the Earth spread over a a distance that is
\textit{significantly} larger than the Solar System (the mean distance between
Pluto and the Sun is $\sim 2\times 10^{-4}\,{\rm pc}$). Even the lowest-mass,
earliest-collapsing CDM structures are incredibly diffuse compared to typical
astrophysical objects. Although there may be $\mathcal{O} (10^{17})$ Earth-mass
dark matter subhalos within the Milky Way's $\approx 300\,\kpc$ virial radius,
detecting them is a daunting challenge.
\end{textbox}

\section{Summary and Outlook}
\label{sec:outlook}
Small-scale structure sits at the nexus of astrophysics, particle physics, and
cosmology. Within the standard \lcdm\ model, most properties of small-scale
structure can be modeled with high precision in the limit that baryonic physics
is unimportant. And yet, the level of agreement between theory and observations
remains remarkably hard to assess, in large part because of hard-to-model
effects of baryonic physics on first-principles predictions. Given the stakes --
absent direct detection of dark matter on Earth, indirect evidence from
astrophysics provides the strongest clues to dark matter's nature -- it is
essential to take potential discrepancies seriously and to explore all avenues
for their resolution.

We have discussed three main classes of problems in this review: (1) counts and
(2) densities of low-mass objects, and (3) tight scaling relations between the
dark and luminous components of galaxies. All of these issues may have their
origin in baryonic physics, but they may also point to the need for a
phenomenological theory that goes beyond \lcdm. Understanding which of these two
options is correct is pressing for both astrophysics and particle physics.

In our opinion, the search for abundant dark matter halos with inferred virial masses
substantially lower than the expected threshold of galaxy formation
($\mvir \sim 10^8\,\msun$) is the most urgent calling in this field
today. The existence of these structures is an unambiguous prediction of all
WIMP-based dark matter models (though it is not unique to WIMP models), and
confirmation of the existence of dark matter halos with $M \sim 10^{6}\,\msun$ or
less would strongly constrain particle physics of dark matter and effectively
rule out any role of dark matter free-streaming in galaxy formation. Here, too, accurate
predictions for the number of expected dark subhalos 
will require an honest accounting of baryon physics -- specifically the destructive effects of central
galaxies themselves \citep[e.g.,][]{garrison-kimmel2017a}.  Of nearly
equal importance is characterizing the central dark matter density structure of very faint
($\mstar \la 10^{6}\,\msun$) galaxies, as a prediction of many recent
high-resolution cosmological simulations within the \lcdm\ paradigm is that
stellar feedback from galaxies below this threshold mass should not modify their
host dark matter halos' cuspy density profile shape. The detection of ubiquitous
cores in very low-mass galaxies therefore has the potential to falsify the
\lcdm\ paradigm.

While some of the tests of the paradigm are clear, their implementation is
difficult. Dark matter substructure is extremely diffuse compared to baryonic
matter, making its detection highly challenging. The smallest galaxies have very few
stars to base accurate dynamical studies upon.  Nevertheless, a variety of
independent probes of the small-scale structure of dark matter are now feasible,
and the LSST era will likely provide a watershed for our understanding of the
nature of dark matter and the threshold of galaxy formation. It is not
far-fetched to think that improved astrophysical data, theoretical
understanding, and numerical simulations will provide a definitive test of \lcdm\
within the next decade, even without the direct detection of particle dark
matter on Earth.

\section*{DISCLOSURE STATEMENT}
The authors are not aware of any affiliations, memberships, funding, or
financial holdings that might be perceived as affecting the objectivity of this
review.

\section*{ACKNOWLEDGMENTS}
It is a pleasure to thank our collaborators and colleagues for helpful
discussions and for making important contributions to our perspectives on this
topic. We specifically thank Peter Behroozi, Brandon Bozek, Peter Creasey, Sandy
Faber, Alex Fitts, Shea Garrison-Kimmel, Andrea Macci\`{o}, Stacy McGaugh,
Se-Heon Oh, Manolis Papastergis, Marcel Pawlowski, Victor Robles, Laura Sales,
Eduardo Tollet, Mark Vogelsberger, and Hai-Bo Yu for feedback and help in
preparing the figures. We are also grateful to Ethan Nadler for pointing out
errors in the coefficients in Eqns.~8--10. MBK acknowledges support from The
University of Texas at Austin, from NSF grant AST-1517226, and from NASA grants
HST-AR-12836, HST-AR-13888, HST-AR-13896, and HST-AR-14282 from the Space
Telescope Science Institute (STScI), which is operated by AURA, Inc., under NASA
contract NAS5-26555.  JSB was supported by NSF grant AST-1518291 and by NASA
through HST theory grants (programs AR-13921, AR-13888, and AR-14282) awarded by
STScI.  This work used computational resources granted by the Extreme Science
and Engineering Discovery Environment (XSEDE), which is supported by National
Science Foundation grant number OCI-1053575 and ACI-1053575.  Resources
supporting this work were also provided by the NASA High-End Computing (HEC)
Program through the NASA Advanced Supercomputing (NAS) Division at Ames Research
Center.

\bibliography{araa}
\end{document}

%% file: preamble.tex
\usepackage{color}
\usepackage{bm}
\usepackage{amssymb, aas_macros}
\usepackage{natbib}

\newcommand{\vmax}{V_{\rm{max}}}
\newcommand{\rmax}{R_{\rm{max}}}

\newcommand{\mvir}{M_{\rm{vir}}}

\newcommand{\rvir}{R_{\rm{vir}}}
\newcommand{\vvir}{V_{\rm{vir}}}
\newcommand{\mstar}{M_{\star}}
\newcommand{\mhalo}{M_{\rm halo}}
\newcommand{\msun}{M_{\odot}}
\newcommand{\mpeak}{M_{\rm peak}}

\newcommand{\lsun}{L_{\odot}}
\newcommand{\hmpc}{h^{-1} \, {\rm Mpc}}

\newcommand{\mpc}{{\rm Mpc}}
\newcommand{\kpc}{{\rm kpc}}
\newcommand{\kms}{{\rm km \, s}^{-1}}

\newcommand{\lcdm}{$\Lambda$CDM}
\newcommand{\bmlcdm}{$\bm{\Lambda}$CDM}

\newcommand{\tbtf}{too-big-to-fail}
\newcommand{\gtorder}{\mathrel{\raise.3ex\hbox{$>$}\mkern-14mu
            \lower0.6ex\hbox{$\sim$}}}
\newcommand{\ltorder}{\mathrel{\raise.3ex\hbox{$<$}\mkern-14mu
            \lower0.6ex\hbox{$\sim$}}}